\documentstyle[amssymb,multicol,amsmath]{article}

\newtheorem{Proposition}{Proposition}

\newcommand{\bp}{\begin{Proposition}}
\newcommand{\ep}{\end{Proposition}}
\newcommand{\bi}[1]{\vspace{-3mm} \bibitem{#1}}
\begin{document}

{\Large \it Journal of Physics A {\bf 35} (2002) 5207-5235} \\

\begin{center}
{\Large \bf Quantum Computer with Mixed States
and Four-Valued Logic}
\vskip 5 mm
{\Large \bf Vasily E. Tarasov } \\

{\it Skobeltsyn Institute of Nuclear Physics,
Moscow State University, Moscow 119992, Russia}

{E-mail: tarasov@theory.sinp.msu.ru}
\end{center}

\begin{abstract}
{\small In this paper we discuss a model of quantum computer
in which a state is an operator of density matrix and gates
are general quantum operations, not necessarily unitary.
A mixed state (operator of density matrix) of  n two-level quantum
systems is considered as an element of $4^{n}$-dimensional
operator Hilbert space (Liouville space). It allows to use 
a quantum computer model with four-valued logic. The gates of this
model are general superoperators which act on n-ququat state.
Ququat is a quantum state in a four-dimensional (operator) Hilbert
space. Unitary two-valued logic gates and quantum operations for
an n-qubit open system are considered as four-valued logic gates acting
on n-ququat. We discuss properties of quantum four-valued logic gates.
In the paper we study universality for
quantum four-valued logic gates.
}
\end{abstract}

PACS {03.67.Lx}


\begin{multicols}{2}


\section{Introduction}


The usual models of a quantum computer deal only with unitary gates
on pure states. In these models it is difficult or impossible
to deal formally with measurements, dissipation, decoherence and noise.
It turns out that the restriction to pure states and unitary gates
is unnecessary \cite{AKN,Tarpr,Tarpr2}.
Understanding the dynamics of open systems is important for studying
quantum noise processes \cite{Gar,Schu,SN}, quantum error
correction \cite{Sh,St2,CS,Pr,BDSW}, decoherence effects
in quantum computations \cite{Unr,PSE,BBSS,MPP,LCW,Zan}
and performing simulations of open quantum systems
\cite{Ter,TDV,Bac,LV,AbL,Zal}.

In this paper we generalize the usual model of a quantum computer
to a model in which a state is a density matrix operator
and gates are general superoperators (quantum operations),
not necessarily unitary.
The pure state of n two-level closed quantum systems is an element
of $2^{n}$-dimensional Hilbert space and it allows us to consider
a quantum computer model with two-valued logic. The gates of this
computer model are unitary operators act on a such state.
In the general case, the mixed state (operator of density matrix)
of  n two-level quantum systems is an element of
$4^{n}$-dimensional operator Hilbert space (Liouville space).
It allows us to use a quantum computer model with four-valued logic.
The gates of this model are general superoperators (quantum
operations) which act on general n-ququat state.
A ququat \cite{Tarpr,Tarpr2} is a quantum state in a four-dimensional
(operator) Hilbert space. Unitary gates and quantum operations for
a quantum two-valued logic computer can be considered as four-valued logic
gates of the new model. In the paper we consider universality for
general quantum four-valued logic gates acting on ququats.

In sections II and III  the physical and mathematical backgrounds
(pure and mixed states, Liouville space and superoperators)
are considered.
In section IV, we introduce a generalized computational basis and
generalized computational states for $4^{n}$-dimensional operator
Hilbert space (Liouville space). In section V, we study some
properties of general four-valued logic gates. Unitary gates and
quantum operations of a two-valued logic computer are considered as
four-valued logic gates. In section VI, we introduce a four-valued
classical logic formalism. We realize classical four-valued logic gates
by quantum gates. In section VII, we study universality for
quantum four-valued logic gates. Finally, a short conclusion
is given in section VIII.


\section{Quantum state and qubit}


\subsection{Pure states}

A quantum system in a pure state is described by a unit vector in a
Hilbert space ${\cal H}$. In the Dirac notation a pure state is
denoted by $|\Psi>$. The Hilbert space ${\cal H}$ is a linear
space with an inner product. The inner product  for $ |\Psi_{1}
>$, $|\Psi_{2}> \in {\cal H}$ is denoted by $<\Psi_{1}|\Psi_{2}>$.
A quantum bit or qubit, the fundamental concept of quantum
computations, is a two-state quantum system. The two basis states
labelled $|0>$ and $|1>$ are orthogonal unit vectors, i.e.
\[ <k|l>=\delta_{kl}, \]
where $k,l\in \{0,1\}$.
The Hilbert space of the qubit is ${\cal H}_{2}={\mathbb{C}}^2$.
The quantum system which corresponds to a quantum computer
(quantum circuits) consists of n quantum two-state particles.
The Hilbert space ${\cal H}^{(n)}$ of such a system is a tensor product
of n Hilbert spaces ${\cal H}_{2}$ of one two-state particle: \
${\cal H}^{(n)}={\cal H}_{2}
\otimes {\cal H}_{2} \otimes ... \otimes {\cal H}_{2}$.
The space ${\cal H}^{(n)}$ is a $2^{n}$-dimensional complex linear space.
Let us choose a basis for ${\cal H}^{(n)}$ which consists of the
$2^{n}$ orthonormal states $|k>$, where $k$ is in binary representation.
The state $|k>$ is a tensor product of states $|k_{i}>$ in ${\cal H}^{(n)}$:
\[ |k>=|k_{1}>\otimes |k_{2}> \otimes ... \otimes |k_{n}>=
|k_{1}k_{2}...k_{n}>  , \]
where $k_{i} \in \{0,1\}$ and $i=1,2,...,n$.
This basis is usually called the computational basis
which has $2^{n}$ elements.
A pure state $|\Psi(t)> \in {\cal H}^{(n)}$ is generally a superposition
of the basis states
\begin{equation} \label{Psi}
|\Psi(t)>=\sum^{2^{n}-1}_{k=0}a_{k}(t)|k> ,
\end{equation}
where $\sum^{2^{n}-1}_{k=0} |a_{k}(t)|^{2}=1$.

\subsection{Mixed states}

In general, a quantum system is not in a pure state. Open quantum
systems are not really isolated from the rest of the universe, so
it does not have a well-defined pure state. Landau and von Neumann
introduced a mixed state and a density matrix into quantum theory.
A density matrix is a Hermitian ($\rho^{\dagger}=\rho$), positive
($\rho >0$) operator on ${\cal H}^{(n)}$ with unit trace ($Tr \rho =1$).
Pure states can be characterized by idempotent condition
$\rho^{2}=\rho$.
A pure state (\ref{Psi}) can be represented by the operator
$\rho(t)=|\Psi(t)><\Psi(t)|$.

One can represent an arbitrary density matrix operator $\rho(t)$
for $n$-qubit in terms of tensor products of Pauli
matrices $\sigma_{\mu}$:
\begin{equation}
\label{rhosigma} \rho(t)=\frac{1}{2^{n}} \sum_{\mu_1 ... \mu_n}
P_{\mu_1 ... \mu_n}(t)
\sigma_{\mu_1} \otimes ... \otimes \sigma_{\mu_n} \ . \end{equation}
where $\mu_{i} \in \{0,1,2,3\}$ and $i=1,...,n$.
Here $\sigma_{\mu}$ are Pauli matrices
\begin{equation}
\label{sigma1}
\sigma_{1}=\left(
\begin{array}{cc}
0&1\\
1&0\\
\end{array}
\right), \ \ \
\sigma_{2}=\left(
\begin{array}{cc}
0&-i\\
i&0\\
\end{array}
\right),
\end{equation}
\begin{equation}
\label{sigma2}
\sigma_{3}=\left(
\begin{array}{cc}
1&0\\
0&-1\\
\end{array}
\right), \ \ \
\sigma_{0}=I=\left(
\begin{array}{cc}
1&0\\
0&1\\
\end{array}
\right).
\end{equation}
The real expansion coefficients $P_{\mu_1 ... \mu_n}(t)$ are given by
\[ P_{\mu_1 ... \mu_n}(t)=Tr( \sigma_{\mu_1} \otimes ...
\otimes \sigma_{\mu_n} \rho(t)). \] Normalization ($Tr \rho=1$)
requires that $P_{0...0}(t)=1$. Since the eigenvalues of the Pauli
matrices are $\pm 1$, the expansion coefficients satisfy
$|P_{\mu_1...\mu_n}(t)|\le 1$.


\section{Liouville space and superoperators}


For the concept of Liouville space and superoperators, see
\cite{Cra}-\cite{kn2}.

\subsection{Liouville space}

The space of linear operators acting on a $2^{n}$-dimensional Hilbert
space ${\cal H}^{(n)}$ is a $(2^{n})^{2}=4^{n}$-dimensional complex
linear space $\overline {\cal H}^{(n)}$.
We denote an element $A$ of $\overline{\cal H}^{(n)}$
by a ket-vector $|A)$. The inner product of two elements $|A)$ and
$|B)$ of $\overline{\cal H}^{(n)}$ is defined as
\begin{equation} \label{inner} (A|B)=Tr(A^{\dagger} B) . \end{equation}
The norm $\|A\|=\sqrt{(A|A)}$ is the Hilbert-Schmidt norm of
operator $A$. A new Hilbert space $\overline{\cal H}^{(n)}$ with scalar
product (\ref{inner}) is called the Liouville space attached
to ${\cal H}^{(n)}$ or the associated Hilbert space,
or Hilbert-Schmidt space \cite{Cra}-\cite{kn2}.

Let $\{|k>\}$ be an orthonormal basis of ${\cal H}^{(n)}$:
\[ <k|k'>=\delta_{kk'} \ , \quad \sum^{2^{n}-1}_{k=0}|k><k|=I. \]
Then $|k,l)=||k><l|)$
is an orthonormal basis of the Liouville space $\overline{\cal H}^{(n)}$:
\begin{equation} \label{onb} (k,l|k',l')=\delta_{kk'}\delta_{ll'} , \quad
\sum^{2^{n}-1}_{k=0} \sum^{2^{n}-1}_{l=0}|k,l)(k,l|=\hat I . \end{equation}
This operator basis has $4^{n}$ elements. Note that
\begin{equation} \label{k,l}
|k,l)=|k_{1},l_{1})\otimes |k_{2},l_{2}) \otimes ...
\otimes |k_{n},l_{n}) , \end{equation}
where $k_{i},l_{i} \in \{0,1\}$, $i=1,...,n$, and
\[ |k_{i},l_{i})\otimes |k_{j},l_{j})=
| \ |k_{i}>\otimes  |k_{j}>, <l_{i}| \otimes <l_{j}| \ ). \]
For an arbitrary element $|A)$ of $\overline{\cal H}^{(n)}$ we have
\begin{equation} \label{|A)}
|A)=\sum^{2^{n}-1}_{k=0}\sum^{2^{n}-1}_{l=0} |k,l)(k,l|A) \end{equation}
with
\[ (k,l|A)= Tr( |l><k| A )=<k|A|l>=A_{kl}. \]

\subsection{Superoperators}

Operators which act on $\overline{\cal H}^{(n)}$ are called
superoperators and we denote them in general by the hat.

For an arbitrary superoperator $\hat {\cal E}$ on
$\overline{\cal H}^{(n)}$ we have
\[ (k,l|\hat {\cal E}|A)=\sum^{2^{n}-1}_{k'=0}\sum^{2^{n}-1}_{l'=0}
(k,l|\hat {\cal E}|k',l') (k',l'|A)=\]
\[ =\sum^{2^{n}-1}_{k'=0}\sum^{2^{n}-1}_{l'=0}
{\cal E}_{klk'l'}A_{k'l'}. \]

Let $A$ be a linear operator in Hilbert space ${\cal H}^{(n)}$.
Then the superoperators $\hat L_{A}$ and $\hat R_{A}$ will be defined by
\begin{equation} \label{LR}
\hat L_{A}|B)=|AB) , \quad \hat R_{A}|B)=|BA).
\end{equation}

The superoperator $\hat {\cal P}=|A)(B|$ is defined by
\begin{equation} \label{calP}
\hat {\cal P}|C)=|A)(B|C)=|A)Tr(B^{\dagger}C). \end{equation}

The superoperator $\hat{\cal E}^{\dagger}$ is called the adjoint
superoperator for $\hat{\cal E}$ if
\begin{equation} (\hat{\cal E}^{\dagger}(A)|B)=(A|\hat{\cal E}(B))
\end{equation}
for all $|A)$ and $|B)$ from $\overline{\cal H}^{(n)}$.
For example, if $\hat{\cal E}=\hat L_{A}\hat R_{B}$, then
$\hat{\cal E}^{\dagger}=\hat L_{A^{\dagger}}\hat R_{B^{\dagger}}$.
If $\hat{\cal E}=\hat L_{A}$, then
$\hat{\cal E}^{\dagger}=\hat L_{A^{\dagger}}$.

A superoperator $\hat{\cal E}$ is called unital if $\hat{\cal E}|I)=|I)$.


\section{Generalized computational basis and ququats}


Let us introduce a generalized computational basis and generalized
computational  states for $4^{n}$-dimensional operator Hilbert space
(Liouville space).

\subsection{Pauli representation}

Pauli matrices (\ref{sigma1}) and (\ref{sigma2}) can be considered as 
a basis in operator space. Let us write the Pauli matrices 
(\ref{sigma1}) and (\ref{sigma2}) in the form
\[ \sigma_{1}=|0><1|+|1><0|=|0,1)+|1,0), \]
\[ \sigma_{2}=-i|0><1|+i|1><0|=-i(|0,1)-|1,0) ), \]
\[ \sigma_{3}=|0><0|-|1><1|=|0,0)-|1,1), \]
\[ \sigma_{0}=I=|0><0|+|1><1|=|0,0)+|1,1). \]
Let us use the formulae
\[ |0,0)=\frac{1}{2}(|\sigma_{0})+|\sigma_{3})) \ , \quad
|1,1)=\frac{1}{2}(|\sigma_{0})-|\sigma_{3})), \]
\[ |0,1)=\frac{1}{2}(|\sigma_{1})+i|\sigma_{2})) \ , \quad
|1,0)=\frac{1}{2}(|\sigma_{1})-i|\sigma_{2})). \]
It allows us to rewrite the operator basis
\[ |k,l)=|k_{1},l_{1})\otimes |k_{2},l_{2}) \otimes ...
\otimes |k_{n},l_{n}) \]
by complete basis operators
\[ |\sigma_{\mu})=|\sigma_{\mu_1} \otimes \sigma_{\mu_2} \otimes ...
\otimes \sigma_{\mu_n}), \]
where $\mu_i=2k_{i}+l_{i}$, i.e. $\mu_{i} \in \{0,1,2,3\}$ and $i=1,...,n$.
The basis $|\sigma_{\mu})$ is orthogonal
\[ (\sigma_{\mu}|\sigma_{\mu '})=2^{n} \delta_{\mu \mu '} \]
and complete operator basis
\[ \frac{1}{2^{n}} \sum^{N-1}_{\mu}
|\sigma_{\mu})(\sigma_{\mu}|=\hat I, \]
where $N=4^{n}$. For an arbitrary element $|A)$ of
$\overline{\cal H}^{(n)}$ we have the Pauli representation by
\begin{equation} |A)=\frac{1}{2^{n}}\sum^{N-1}_{\mu=0}
|\sigma_{\mu})(\sigma_{\mu}|A) \end{equation}
with the complex coefficients
$(\sigma_{\mu}|A)=Tr( \sigma_{\mu} A )$.
We can rewrite formula (\ref{rhosigma}) using the complete
operator basis $|\sigma_{\mu})$ in Liouville space
$\overline{\cal H}^{(n)}$:
\begin{equation} |\rho(t))=\frac{1}{2^n}\sum^{N-1}_{\mu=0}
|\sigma_{\mu}) P_{\mu}(t), \end{equation}
where $\sigma_{\mu}=\sigma_{\mu_1} \otimes ... \otimes \sigma_{\mu_n}$,
\ $\mu=(\mu_1...\mu_n)$, $N=4^{n}$ and $P_{\mu}(t)=(\sigma_{\mu}|\rho(t))$.

The density matrix operator $\rho(t)$ is a self-adjoint operator with
unit trace. It follows that
\[ P^{*}_{\mu}(t)=P_{\mu}(t) , \quad P_{0}(t)=(\sigma_{0}|\rho(t))=1. \]
In the general case,
\begin{equation} \label{le} \frac{1}{2^n}
\sum^{N-1}_{\mu=0} P^{2}_{\mu}(t)=
(\rho(t)|\rho(t))=Tr(\rho^{2}(t)) \le 1. \end{equation}
Note that the Schwarz inequality
\[ |(A|B)|^{2} \le (A|A)(B|B)  \]
leads to
\[ |(I|\rho(t))|^{2} \le (I|I)(\rho(t)|\rho(t)). \]
We rewrite this inequality in the form
\begin{equation} \label{ge} 1=|Tr\rho(t)|^{2} \le
2^{n}Tr(\rho^2(t))=\sum^{N-1}_{\mu=0} P^{2}_{\mu}(t), \end{equation}
where $N=4^{n}$. Using (\ref{le}) and (\ref{ge}) we have
\begin{equation}
\frac{1}{\sqrt{2^{n}}} \le Tr( \rho^{2}(t)) \le 1 \quad or \quad
1 \le \sum^{N-1}_{\mu=0} P^{2}_{\mu}(t) \le 2^{n}. \end{equation}

\subsection{Generalized computational basis}

Let us define the orthonormal basis of Liouville space.
In the general case, the state  $\rho(t)$ of the n-qubit system
is an element of Hilbert space $\overline{\cal H}^{(n)}$.
The basis for $\overline{\cal H}^{(n)}$ consists of the
$2^{2n}=4^n$ orthonormal basis elements denoted by $|\mu)$. \\ 

\noindent {\bf Definition}
{\it The basis of Liouville space $\overline{\cal H}^{(n)}$ is defined by
\begin{equation} \label{mu}
|\mu)=|\mu_1...\mu_n)=\frac{1}{\sqrt{2^{n}}}|\sigma_{\mu})=
\frac{1}{\sqrt{2^{n}}}|\sigma_{\mu_1}
\otimes  ... \otimes \sigma_{\mu_n}), \end{equation}
where $N=4^{n}$, $\mu_i \in \{0,1,2,3\}$ and
\begin{equation} \label{mu2}
(\mu|\mu')=\delta_{\mu \mu'} \ , \quad
\sum^{N-1}_{\mu=0} |\mu)(\mu|=\hat I, \end{equation}
is called the "generalized computational basis".} \\ 

Here $\mu$ is a four-valued representation of
\begin{equation} \mu=\mu_1 4^{n-1}+...+\mu_{n-1}4+\mu_n  .
\end{equation}

The pure state of n two-level closed quantum systems is an element
of $2^{n}$-dimensional functional Hilbert space ${\cal H}^{(n)}$.
It leads to a quantum computer model with two-valued logic.
{\it In the general case, the mixed state $\rho(t)$ of $n$ two-level
(open or closed) quantum system is an element of $4^{n}$-dimensional
operator Hilbert space $\overline{\cal H}^{(n)}$ (Liouville space).
It leads to a four-valued logic model for the quantum computer.}

The state $|\rho(t))$ of the quantum computation at any point in time
is a superposition of basis elements, 
\begin{equation} \label{rho3}
|\rho(t))=\sum^{N-1}_{\mu=0} |\mu)\rho_{\mu}(t),
\end{equation}
where $\rho_{\mu}(t)=(\mu|\rho(t))$ are real numbers (functions)
satisfying normalized condition 
\begin{equation} \label{rho-nc} 
\rho_0(t)=\frac{1}{\sqrt{2^n}}(\sigma_0|\rho(t))=
\frac{1}{\sqrt{2^n}}Tr(\rho(t))=\frac{1}{\sqrt{2^n}}. \end{equation}

\subsection{Generalized computational states}

Generalized computational basis elements $|\mu)$ are not quantum states
for $\mu\not=0$. It follows from normalized condition
(\ref{rho-nc}).
The general quantum state in the Pauli representation 
has the form (\ref{rho3}).
Let us define simple computational quantum states. \\

\noindent {\bf Definition}
{\it Quantum states in Liouville space defined by
\begin{equation} |\mu]=\frac{1}{2^{n}}\Bigl(|\sigma_{0})+
|\sigma_{\mu}) (1-\delta_{\mu 0}) \Bigr) \end{equation}
or
\begin{equation} |\mu]=\frac{1}{\sqrt{2^{n}}}\Bigl(|0)+
|\mu)(1-\delta_{\mu 0}) \Bigr). \end{equation}
are called "generalized computational states".} \\

Note that all states $|\mu]$, where $\mu \not=0$, are pure states,
since $[\mu|\mu]=1$. The state $|0]$ is a maximally mixed state.
The states $|\mu]$ are elements of Liouville space
$\overline{\cal H}^{(n)}$.

The quantum state in a four-dimensional Hilbert space is usually called
ququat, qu-quart \cite{BPT} or qudit \cite{Run,CBKG,DW,BB,CMP} with $d=4$.
Usually the ququat is considered as a four-level quantum system.
We consider the ququat as a general quantum state in a four-dimensional
operator Hilbert space. \\

\noindent {\bf Definition}
{\it A quantum state in four-dimensional operator Hilbert space
(Liouville space)
$\overline{\cal H}^{(1)}$ associated with a single qubit of
space ${\cal H}^{(1)}={\cal H}_2$ is called a "single ququat".
A quantum state in $4^n$-dimensional Liouville space
$\overline{\cal H}^{(n)}$ associated with an n-qubit system
is called an "n-ququat".} \\

\noindent
{\bf Example.}
For the single ququat the states $|\mu]$ are
\[ |0]=\frac{1}{2}|\sigma_0) , \quad
|k]=\frac{1}{2}\Bigl(|\sigma_{0})+|\sigma_{k})\Bigr) , \]
or
\[ |0]=\frac{1}{\sqrt{2}}|0) , \quad
|k]=\frac{1}{\sqrt{2}}\Bigl(|0)+|k)\Bigr). \]

It is convenient to use matrices for quantum states.
In matrix representation the single ququat computational basis
$|\mu)$ can be represented by

$$|0)=\left(
\begin{array}{c}
1\\
0\\
0\\
0
\end{array}
\right),
|1)=\left(
\begin{array}{c}
0\\
1\\
0\\
0
\end{array}
\right),
|2)=\left(
\begin{array}{c}
0\\
0\\
1\\
0
\end{array}
\right),
|3)=\left(
\begin{array}{c}
0\\
0\\
0\\

1
\end{array}
\right).
$$

In this representation single ququat generalized computational
states $|\mu]$ are represented by
$$|0]=\frac{1}{\sqrt{2}}\left(
\begin{array}{c}
1\\
0\\
0\\
0
\end{array}
\right),
\quad
|1]=\frac{1}{\sqrt{2}}\left(
\begin{array}{c}
1\\
1\\
0\\
0
\end{array}
\right),$$
$$|2]=\frac{1}{\sqrt{2}}\left(
\begin{array}{c}
1\\
0\\
1\\
0
\end{array}
\right),
\quad
|3]=\frac{1}{\sqrt{2}}\left(
\begin{array}{c}
1\\
0\\
0\\
1
\end{array}
\right).
$$
A general single ququat quantum state
$|\rho)=\sum^{3}_{\mu=0}|\mu)\rho_{\mu}$ is represented by
$$|\rho)=\left(
\begin{array}{c}
\rho_{0}\\
\rho_{1}\\
\rho_{2}\\
\rho_{3}
\end{array}
\right),
$$
where $\rho_{0}=1/\sqrt{2}$ and
$\rho^{2}_{1}+\rho^{2}_{2}+\rho^{2}_{3} \le \sqrt{2}$.

We can use the other matrix representation for the states
$|\rho]$ which have no  coefficient $1/\sqrt{2^{n}}$.
In this representation single ququat generalized computational
states $|\mu]$ are represented by
$$|0]=\left[
\begin{array}{c}
1\\
0\\
0\\
0
\end{array}
\right],\
|1]=\left[
\begin{array}{c}
1\\
1\\
0\\
0
\end{array}
\right],\
|2]=\left[
\begin{array}{c}
1\\
0\\
1\\
0
\end{array}
\right],\
|3]=\left[
\begin{array}{c}
1\\
0\\
0\\
1
\end{array}
\right].
$$
A general single ququat quantum state
$$|\rho]=\left[
\begin{array}{c}
1\\
P_{1}\\
P_{2}\\
P_{3}
\end{array}
\right],
$$
where $P^{2}_{1}+P^{2}_{2}+P^{2}_{3} \le 1$, is a superposition
of generalized computational states
\[ |\rho]=|0](1-P_1-P_2-P_3)+|1] P_1+|2]P_2+|3] P_3. \]


\section{Quantum four-valued logic gates}


\subsection{Superoperators and quantum gates}

Unitary evolution is not the most general type
of state change possible for quantum systems.
The most general state change of a quantum system is a positive map
which is called a quantum operation or superoperator.
For the concept of quantum operations, see \cite{Kr1,Kr2,Kr3,Kr4,Schu}.

Quantum operations can be considered as generalized quantum gates
acting on general (mixed) states.
Let us define a quantum four-valued logic gate. \\

\noindent {\bf Definition}
{\it A quantum four-valued logic gate is a superoperator $\hat{\cal E}$
on Liouville space $\overline{\cal H}^{(n)}$ which maps a density
matrix operator $|\rho)$ of $n$-ququat to a density matrix operator
$|\rho')$ of $n$-ququat.} \\

Let us consider a superoperator $\hat{\cal E}$ which
maps density matrix operator $|\rho)$ to density matrix operator
$|\rho^{\prime})$.
If $|\rho)$ is a density matrix operator, then $\hat{\cal E}|\rho)$
should also be a density matrix operator.
Therefore we have some requirements for superoperator $\hat{\cal E}$.
The requirements for a superoperator $\hat{\cal E}$ on the space 
${\cal H}^{(n)}$ to be the quantum four-valued logic gate are as follows: \\
\begin{enumerate}
\item The superoperator $\hat{\cal E}$ is a {\it real} superoperator, i.e.
$\Bigl(\hat {\cal E}(A)\Bigr)^{\dagger}=\hat {\cal E}(A^{\dagger})$
for all $A$ or
$\Bigl(\hat {\cal E}(\rho)\Bigr)^{\dagger}=\hat {\cal E}(\rho)$.
The real superoperator $\hat{\cal E}$ maps self-adjoint operator
$\rho$ to self-adjoint operator $\hat{\cal E}(\rho)$:
$ (\hat{\cal E}(\rho))^{\dagger}=\hat{\cal E}(\rho)$.
\item The superoperator $\hat{\cal E}$ is a {\it positive} superoperator,
i.e. $\hat{\cal E}$ maps positive operators to positive operators: \
$\hat {\cal E}(A^{2}) >0$ for all $A\not=0$
or $\hat{\cal E}(\rho)\ge 0$.

We have to assume the superoperator $\hat{\cal E}$
to be not merely positive but completely positive. The
superoperator $\hat{\cal E}$ is a {\it completely positive} map of
Liouville space, i.e. the positivity remains if we extend the
Liouville space $\overline{\cal H}^{(n)}$ by adding more qubits.
That is, the superoperator $\hat{\cal E} \otimes \hat I^{(m)}$
must be positive, where $\hat I^{(m)}$ is the identity
superoperator on some Liouville space $\overline{\cal H}^{(m)}$.
\item The superoperator $\hat{\cal E}$ is a {\it trace-preserving} map, i.e.
\begin{equation} \label{EI}
(I|\hat{\cal E}|\rho)=(\hat{\cal E}^{\dagger}(I)|\rho)=1 \quad
or \quad \hat{\cal E}^{\dagger}(I)=I. \end{equation}
\end{enumerate}

As the result we have the following definition. \\

\noindent {\bf Definition}
{\it Quantum four-valued logic gate is
a real positive (or completely positive) trace-preserving
superoperator $\hat{\cal E}$
on Liouville space $\overline{\cal H}^{(n)}$.} \\

In the general case, we can consider linear and nonlinear quantum
four-valued logic gates.
Let the superoperator $\hat{\cal E}$ be a {\it convex linear} map
on the set of density matrix operators, i.e.
\[ \hat{\cal E}(\sum_{s} \lambda_{s} \rho_{s})=
\sum_{s} \lambda_{s} \hat{\cal E}(\rho_{s}), \]
where all $\lambda_{s}$ are $0<\lambda_{s}<1$ and
$\sum_{s} \lambda_{s}=1$.
Any convex linear map of density matrix operators
can be uniquely extended to a {\it linear} map on Hermitian operators.
Note that any linear completely positive superoperator can be represented by
\[ \hat{\cal E}=\sum^{m}_{j=1}\hat L_{A_{j}}\hat R_{A^{\dagger}_{j}}. \]
If this superoperator is a trace-preserving superoperator, then
\[ \sum^m_{j=1} A^{\dagger}_j A_j=I , \]
i.e. the condition (\ref{EI}) is satisfied.

The restriction to linear gates is unnecessary.
Let us consider a linear real completely positive
superoperator $\hat {\cal E}$ which is not
trace-preserving. This superoperator is not a quantum gate.
Let $(I|\hat{\cal E}|\rho)=Tr(\hat{\cal E}(\rho))$
be the probability that the process represented
by the superoperator $\hat{\cal E}$ occurs.
Since the probability is non-negative and
never exceed 1, it follows that the superoperator
$\hat{\cal E}$ is a trace-decreasing superoperator:
\[ 0\le (I|\hat{\cal E}|\rho) \le 1 \quad
or \quad \hat{\cal E}^{\dagger}(I) \le I. \]
In general, any linear real completely positive
trace-decreasing superoperator generates
a quantum four-valued logic gate.
The quantum four-valued logic gate can be defined as {\it nonlinear}
superoperator $\hat{\cal N}$ by
\begin{equation} \label{Ngate} \hat{\cal N}|\rho)=
(I|\hat{\cal E}|\rho)^{-1} \hat{\cal E}|\rho) \quad or
\quad \hat{\cal N}(\rho)=
\frac{\hat{\cal E}(\rho)}{Tr(\hat{\cal E}(\rho))}, \end{equation}
where $\hat{\cal E}$ is a linear real completely positive
trace-decreasing superoperator.

In the generalized computational basis the gate
$\hat{\cal E}$ can be represented by
\begin{equation} \hat{\cal E}=\frac{1}{2^n}\sum^{N-1}_{\mu=0}
\sum^{N-1}_{\nu=0}
{\cal E}_{\mu \nu} |\sigma_{\mu})(\sigma_{\nu}|. \end{equation}
where $N=4^n$, $\mu$ and $\nu$ are four-valued representations of
\[ \mu=\mu_1 4^{N-1}+...+\mu_{N-1}4+\mu_N , \]
\[ \nu=\nu_1 4^{N-1}+...+\nu_{N-1}4+\nu_N, \]
\[ \sigma_{\mu}=\sigma_{\mu_1} \otimes...\otimes \sigma_{\mu_n}, \]
$\mu_i,\nu_i \in \{0,1,2,3\}$ and
${\cal E}_{\mu \nu}$ are elements of some matrix.

\subsection{General quantum operation as four-valued logic gates}

Unitary gates and quantum operations for a quantum computer
with pure states and two-valued logic can be considered
as quantum four-valued logic gates acting on mixed states. 

\bp
{\it In the generalized computational basis $|\mu)$ any linear
quantum operation $\hat {\cal E}$ acting on n-qubit mixed (or pure)
states can be represented as a quantum four-valued logic gate
$\hat{\cal E}$ on n-ququat by
\begin{equation} \hat{\cal E}=\sum^{N-1}_{\mu=0}\sum^{N-1}_{\nu=0}
{\cal E}_{\mu \nu} \ |\mu)(\nu| , \end{equation}
where $N=4^{n}$,
\begin{equation} {\cal E}_{\mu \nu}=\frac{1}{2^n}
Tr\Bigl(\sigma_{\mu} \hat {\cal E} (\sigma_{\nu}) \Bigr), \end{equation}
and  $\sigma_{\mu} =\sigma_{\mu_1} \otimes ...  \otimes \sigma_{\mu_n}$.}
\ep

\noindent {\bf Proof.}
The state $\rho^{\prime}$ in the generalized computational basis
$|\mu)$ has the form
\[ |\rho^{\prime})=\sum^{N-1}_{\mu=0} |\mu)\rho^{\prime}_{\mu} \ , \]
where $N=4^{n}$ and
\[ \rho^{\prime}_{\mu}=(\mu|\rho^{\prime})=
\frac{1}{\sqrt{2^n}}Tr(\sigma_{\mu}\rho^{\prime}). \]
The quantum operation $\hat {\cal E}$ defines a quantum
four-valued logic gate by
\[ |\rho^{\prime})= \hat{\cal E}|\rho) =|\hat {\cal E}(\rho) )=
\sum^{N-1}_{\nu=0} |\hat {\cal E}(\sigma_{\nu}) )
\frac{1}{\sqrt{2^n}} \rho_{\nu}. \]
Then
\[ (\mu|\rho^{\prime})=\sum^{N-1}_{\nu =0}
(\sigma_{\mu}|\hat {\cal E}( \sigma_{\nu}) )
\frac{1}{2^n} \rho_{\nu}. \]
Finally, we obtain
\[ \rho^{\prime}_{\mu}=\sum^{N-1}_{\nu=0}
{\cal E}_{\mu \nu}\rho_{\nu}, \]
where
\[ {\cal E}_{\mu \nu}=\frac{1}{2^n}
(\sigma_{\mu}|\hat {\cal E}(\sigma_{\nu}) )
=\frac{1}{2^n}Tr\Bigl(\sigma_{\mu} \hat {\cal E} (\sigma_{\nu}) \Bigr). \]
This formula defines a relation between quantum
operation $\hat {\cal E}$ and the real $4^{n}\times 4^{n}$ matrix
${\cal E}_{\mu \nu}$ of a quantum four-valued logic gate. \ \ \ $\Box$

Quantum four-valued logic gates $\hat{\cal E}$
in the matrix representation are represented by $4^{n}\times 4^{n}$
matrices ${\cal E}_{\mu \nu}$.
The matrix of the gate $\hat{\cal E}$ is
$${\cal E}=\left(
\begin{array}{cccc}
{\cal E}_{00}&{\cal E}_{01}&...&{\cal E}_{0a}\\
{\cal E}_{10}&{\cal E}_{11}&...&{\cal E}_{1a}\\
...&...& ...&...\\
{\cal E}_{a0}&{\cal E}_{a1}&...
&{\cal E}_{aa}
\end{array}
\right),$$
where $a=N-1=4^{n}-1$.
In matrix representation the gate $\hat{\cal E}$ maps
the state $|\rho)=\sum^{N-1}_{\nu=0}|\nu) \rho_{\nu}$ to
the state $|\rho^{\prime})=\sum^{N-1}_{\mu=0}|\mu) \rho^{\prime}_{\mu}$ by
\begin{equation} \label{rEr} \rho^{\prime}_{\mu}=\sum^{N-1}_{\nu=0}
{\cal E}_{\mu \nu} \rho_{\nu} , \end{equation}
where $\rho^{\prime}_{0}=\rho_{0}=1/\sqrt{2^{n}}$.
It can be written in the form
$$\left(
\begin{array}{c}
\rho^{\prime}_0\\
\rho^{\prime}_1\\
...\\
\rho^{\prime}_{a}
\end{array}
\right)=
\left(
\begin{array}{ccccc}
{\cal E}_{00}&{\cal E}_{01}&...&{\cal E}_{0a}\\
{\cal E}_{10}&{\cal E}_{11}&...&{\cal E}_{1a}\\
...&...& ...&...\\
{\cal E}_{a0}&{\cal E}_{a1}&...
&{\cal E}_{aa}
\end{array}
\right)
\left(
\begin{array}{c}
\rho_0\\
\rho_1\\
...\\
\rho_{a}
\end{array}
\right).$$

Since $P_{\mu}=\sqrt{2^{n}} \rho_{\mu}$ and
$P^{\prime} _{\mu}=\sqrt{2^{n}} \rho^{\prime}_{\mu}$,
it follows that representation (\ref{rEr}) for linear gate
$\hat{\cal E}$ is equivalent to
\begin{equation} \label{PEP} P^{\prime}_{\mu}=\sum^{N-1}_{\nu=0}
{\cal E}_{\mu \nu} P_{\nu} . \end{equation}
It can be written in the form
$$\left[
\begin{array}{c}
P^{\prime}_0\\
P{\prime}_1\\
...\\
P^{\prime}_{a}
\end{array}
\right]=
\left(
\begin{array}{cccc}
{\cal E}_{00}&{\cal E}_{01}&...&{\cal E}_{0a}\\
{\cal E}_{10}&{\cal E}_{11}&...&{\cal E}_{1a}\\
...&...& ...&...\\
{\cal E}_{a0}&{\cal E}_{a1}&...
&{\cal E}_{aa}
\end{array}
\right)
\left[
\begin{array}{c}
P_0\\
P_1\\
...\\
P_{a}
\end{array}
\right].$$
where $P_{0}=P^{\prime}_0=1$.
Note that if we use different matrix representations of state
we can use identical matrices for gate $\hat{\cal E}$. \ \ \ $\Box$

\bp
{\it In the generalized computational basis $|\mu)$ the matrix
${\cal E}_{\mu \nu}$ of a linear quantum four-valued
logic gate
\begin{equation}  \label{elr}
\hat{\cal E}=\sum^{m}_{j=1} \hat L_{A_j} \hat R_{A^{\dagger}_{j}}
\end{equation}
is real, i.e.
${\cal E}^{*}_{\mu \nu}={\cal E}_{\mu \nu}$ for all
$\mu$ and $\nu$.
Any real matrix ${\cal E}_{\mu \nu}$ associated with
linear (trace-preserving) quantum four-valued
logic gate (\ref{elr}) has
\begin{equation} {\cal E}_{0 \nu}=\delta_{0 \nu}. \end{equation}
}
\ep

\noindent {\bf Proof.}
\[ {\cal E}_{\mu \nu}=\frac{1}{2^n}\sum^{m}_{j=1}
Tr\Bigl(\sigma_{\mu} A_{j} \sigma_{\nu} A^{\dagger}_{j} \Bigr)=
\frac{1}{2^n} \sum^{m}_{j=1}
(A^{\dagger}_j  \sigma_{\mu}| \sigma_{\nu}A^{\dagger}_j ). \]
\[ {\cal E}^{*}_{\mu \nu}=\frac{1}{2^n} \sum^{m}_{j=1}
(A^{\dagger}_j \sigma_{\mu}| \sigma_{\nu} A^{\dagger}_j)^{*}=
\frac{1}{2^n} \sum^{m}_{j=1}
( \sigma_{\nu} A^{\dagger}_j|A^{\dagger}_j  \sigma_{\mu} )=  \]
\[ =\frac{1}{2^n} \sum^{m}_{j=1}
Tr\Bigl(A_j \sigma_{\nu} A^{\dagger}_{j} \sigma_{\mu} \Bigr)=
\frac{1}{2^n} \sum^{m}_{j=1}
Tr\Bigl( \sigma_{\mu} A_{j} \sigma_{\nu} A^{\dagger}_j \Bigr)=
{\cal E}_{\mu \nu}. \]

Let us consider the ${\cal E}_{0 \nu}$ for gate (\ref{elr}):
\[ {\cal E}_{0 \nu}=\frac{1}{2^{n}}
Tr\Bigl(\sigma_{0}{\cal E}(\sigma_{\nu}) \Bigr)=
\frac{1}{2^{n}} Tr\Bigl({\cal E}(\sigma_{\nu}) \Bigr)= \]
\[ =\frac{1}{2^{n}}
Tr\Bigl(\sum^{m}_{j=1} A_j \sigma_{\nu}A^{\dagger}_j \Bigr)=
\frac{1}{2^{n}} Tr\Bigl((\sum^{m}_{j=1}
A^{\dagger}_j A_j )\sigma_{\nu} \Bigr)=\]
\[=\frac{1}{2^{n}} Tr \sigma_{\nu}=\delta_{0\nu}. \]

In the general case, a linear quantum four-valued logic gate acts on $|0)$ by
\[ \hat{\cal E}|0)=|0)+\sum^{N-1}_{k=1}T_{k}|k). \]
For example, a single ququat quantum gate acts by
\[ \hat{\cal E}|0)=|0)+T_{1}|1)+T_{2}|2)+T_{3}|3). \]
If all $T_{k}$, where $k=1,...,N-1$ are equal to zero,
then $\hat{\cal E}|0)=|0)$. The linear quantum gates with $T=0$
conserve the maximally mixed state $|0]$ invariant. \\

\noindent {\bf Definition} {\it A quantum four-valued logic gate
$\hat{\cal E}$ is called a unital gate or gate with $T=0$ if the 
maximally mixed state $|0]$ is invariant under the action of this
gate: $\hat{\cal E}|0]=|0]$.} \\

The output state of a linear quantum four-valued logic gate
$\hat{\cal E}$ is $|0]$
if and only if the input state is $|0]$.
If $\hat{\cal E}|0]\not=|0]$, then
$\hat{\cal E}$ is not a unital gate.

\bp
{\it The matrix ${\cal E}_{\mu \nu}$ of linear real trace-preserving
superoperator $\hat{\cal E}$ on n-ququat is an element
of group $TGL(4^n-1,\mathbb{R})$
which is a semidirect product of general linear group
$GL(4^n-1,\mathbb{R})$ and translation group $T(4^n-1,\mathbb{R})$.}
\ep

\noindent {\bf Proof.}
This proposition follows from proposition 2.
Any element (matrix ${\cal E}_{\mu \nu}$) of group $TGL(4^n-1,\mathbb{R})$
can be represented by
$${\cal E}(T,R)=\left(
\begin{array}{cc}
1&0\\
T&R
\end{array}
\right),$$
where $T$ is a column with $4^n-1$ elements,
$0$ is a line with $4^n-1$ zero elements, and
$R$ is a real $(4^n-1)\times (4^n-1)$ matrix
$R \in GL(4^n-1,\mathbb{R})$.
If $R$ is orthogonal $(4^{n}-1)\times(4^{n}-1)$ matrix ($R^TR=I$),
then we have the motion group \cite{Vil,Vil1,Or}.
The group multiplication of elements ${\cal E}(T,R)$ and ${\cal E}(T',R')$
is defined by
\[ {\cal E}(T,R){\cal E}(T',R')={\cal E}(T+RT',RR'). \]
In particular, we have
\[ {\cal E}(T,R)={\cal E}(T,I){\cal E}(0,R) \ , \quad
{\cal E}(T,R)={\cal E}(0,R){\cal E}(R^{-1}T,I). \]
where $I$ is a unit $(4^n-1)\times (4^n-1)$ matrix.

Any linear real trace-preserving superoperator can be decompose 
into unital superoperator and translation superoperator.
It allows us to consider two types of linear trace-preserving superoperators:\\
(1) Unital superoperators $\hat{\cal E}^{(T=0)}$ with
the matrices ${\cal E}(0,R)$.
The n-ququat unital superoperator can be represented by
\[ \hat{\cal E}^{(T=0)}=|0)(0|+\sum^{N-1}_{k=1}
\sum^{N-1}_{l=1} R_{kl}|k)(l|, \]
where $N=4^{n}$.\\
(2) Translation superoperators $\hat{\cal E}^{(T)}$
defined by matrices ${\cal E}(T,I)$ and
\[ \hat{\cal E}^{(T)}=\sum^{N-1}_{\mu=0} |\mu)(\mu|+
\sum^{N-1}_{k=1}T_k |k)(0|. \]

\subsection{Decomposition for linear superoperators}

Let us consider the n-ququat linear real superoperator
\begin{equation} \label{LGE}
\hat{\cal E}=|0)(0|+\sum^{N-1}_{\mu=1} T_{\mu} |\mu)(0|+
\sum^{N-1}_{\mu=1} \sum^{N-1}_{\nu=1} R_{\mu \nu} |\mu)(\nu|,
\end{equation}
where $N=4^{n}$.

\bp (Singular valued decomposition for matrix)\\
{\it Any real  matrix $R$ can be written in the form
$R={\cal U}_{1} D {\cal U}^{\small T}_{2},$
where
${\cal U}_{1}$ and ${\cal U}_{2}$  are
real orthogonal $(N-1)\times(N-1)$ matrices and
$D=diag(\lambda_1,...,\lambda_{N-1})$
is a diagonal $(N-1)\times(N-1)$  matrix
such that \ \ 
$\lambda_1 \ge \lambda_2 \ge ... \ge \lambda_{N-1} \ge 0$.

}
\ep

\noindent {\bf Proof.}
This proposition is proved  in \cite{EY,Lan,Sc,Gant}. \ \ \ $\Box$ \\

In the general case, we have the following proposition.

\bp
(Singular valued decomposition for superoperator)\\
{\it Any linear real superoperator (\ref{LGE})
can be represented by
\begin{equation} \hat{\cal E}=\hat{\cal E}^{(T)} \hat{\cal U}_{1} \
\hat D \ \hat{\cal U}_{2}, \end{equation}
where\\
$\hat{\cal U}_{1}$ and $\hat{\cal U}_{2}$ are
unital orthogonal superoperators, such that
\begin{equation} \label{Ui}
\hat{\cal U}_{i}=|0)(0|+\sum^{N-1}_{\mu=1} \sum^{N-1}_{\nu=1}
{\cal U}^{(i)}_{\mu \nu} |\mu)(\nu|,
\end{equation}
$\hat D$ is a unital diagonal superoperator, such that
\begin{equation} \label {D} \hat D=|0)(0|+\sum^{N-1}_{\mu=1}
\lambda_{\mu} |\mu)(\mu|, \end{equation}
where $\lambda_{\mu} \ge 0$.\\
$\hat{\cal E}^{(T)}$ is a translation superoperator, such that
\begin{equation} \hat{\cal E}^{(T)}=|0)(0|+\sum^{N-1}_{\mu=1}
|\mu)(\mu|+\sum^{N-1}_{\mu=1} T_{\mu} |\mu)(0|. \end{equation}
}
\ep

\noindent {\bf Proof.} The proof of this proposition can be easily realized
in matrix representation by using proposition 3 and 4. \ \ \ $\Box$ \\

As a result we have that any linear real trace-preserving
superoperator can be realized by three types of superoperators: \\
(1) unital orthogonal superoperator $\hat{\cal U}$; \\
(2) unital diagonal superoperator $\hat D$; \\
(3) nonunital translation superoperator $\hat{\cal E}^{(T)}$. 


\bp
{\it If the quantum operation $\hat {\cal E}$ has the form
\[ \hat {\cal E}(\rho)=\sum^{m}_{j=1} A_{j} \rho A^{\dagger}_{j}, \]
where $A$ is a self-adjoint operator ($A^{\dagger}_{j}=A_{j}$), then
quantum four-valued logic gate $\hat{\cal E}$ is described
by symmetric matrix ${\cal E}_{\mu \nu}={\cal E}_{\nu \mu}$.
This gate is trace-preserving if
${\cal E}_{\mu 0}={\cal E}_{0 \mu}=\delta_{\mu 0}$.
}
\ep

\noindent {\bf Proof.} If $A^{\dagger}_j=A_j$, then
\[ {\cal E}_{\mu \nu}=
\frac{1}{\sqrt{2^{n}}}\sum^{m}_{j=1}
Tr (\sigma_{\mu}A_{j} \sigma_{\nu} A_{j})=\]
\[=\frac{1}{\sqrt{2^{n}}} \sum^{m}_{j=1}
Tr (\sigma_{\nu}A_{j} \sigma_{\mu} A_{j})={\cal E}_{\nu \mu}. \]
Using proposition 2 we have
that this gate is trace-preserving if
${\cal E}_{\mu 0}={\cal E}_{0 \mu}=\delta_{\mu 0}$. \ \ \ $\Box$

\subsection{Unitary two-valued logic gates as orthogonal
four-valued logic gates}

Let us consider a unitary two-valued logic gate defined
by unitary operator $U$ acting on pure states - unit
elements of Hilbert space.
The map $\hat {\cal U}: \rho \rightarrow U \rho U^{\dagger}$
induced by a unitary operator $U$ is a particular case of
quantum operation.

\bp
{\it In the generalized computational basis any unitary
quantum two-valued logic gate $U$ acts on pure n-qubit states
can be considered as a quantum four-valued logic 
gate $\hat {\cal U}$ acting on n-ququat:
\begin{equation} \label{P6-1}
\hat{\cal U}=\sum^{N-1}_{\mu=0}\sum^{N-1}_{\nu=0}
{\cal U}_{\mu \nu} |\mu)(\nu|  \ , \end{equation}
where ${\cal U}_{\mu \nu}$ is a real matrix such that
\begin{equation} \label{P6-2}
{\cal U}_{\mu \nu}=\frac{1}{2^n}Tr\Bigl(\sigma_{\nu}U
\sigma_{\mu} U^{\dagger}\Bigr) \ . \end{equation}
}
\ep

\noindent {\bf Proof.} Using proposition 1 and the equation
\[ |\rho^{\prime})=\hat{\cal U}|\rho)=
|U \rho U^{\dagger}), \]
we get this proposition. \ \ \ $\Box$ \\

Formulae (\ref{P6-1}) and (\ref{P6-2}) define a relation
between the unitary quantum two-valued logic gate $U$ and
the real $4^{n}\times 4^{n}$ matrix ${\cal U}$ of quantum
four-valued logic gate $\hat {\cal U}$.

\bp
{\it Any four-valued logic gate associated with unitary two-valued
logic gate by (\ref{P6-1}) and (\ref{P6-2}) is a unital gate, i.e.
gate matrix ${\cal U}$ defined by (\ref{P6-2}) has
${\cal U}_{\mu 0}={\cal U}_{0 \mu }=\delta_{\mu 0}$.}
\ep

\noindent {\bf Proof.}
\[ {\cal U}_{\mu 0}=\frac{1}{2^{n}}
Tr\Bigl(\sigma_{\mu}U\sigma_{0} U^{\dagger}\Bigr)=\frac{1}{2^{n}}
Tr\Bigl(\sigma_{\mu}UU^{\dagger}\Bigr)=\frac{1}{2^{n}}
Tr\sigma_{\nu}. \]
Using $Tr\sigma_{\mu}=\delta_{\mu 0}$
we get ${\cal U}_{\mu 0}=\delta_{\mu 0}$. \ \ \ $\Box$ \\

Let us denote the gate $\hat{\cal U}$ associated with unitary
two-valued logic gate $U$ by $\hat{\cal E}^{(U)}$.

\bp
{\it If $U$ is a unitary two-valued logic gate, then
in the generalized computational basis a quantum four-valued
logic gate $\hat{\cal U}=\hat{\cal E}^{(U)}$ associated with $U$
is represented
by orthogonal matrix ${\cal E}^{(U)}$: }
\begin{equation} \label{ORT}
{\cal E}^{(U)}({\cal E}^{(U)})^{T}=
({\cal E}^{(U)})^{T}{\cal E}^{(U)}=I .
\end{equation}
\ep

\noindent {\bf Proof.}
Let $\hat{\cal E}^{(U)}$ be defined by
\[ \hat{\cal E}^{(U)}|\rho)=|U \rho U^{\dagger}) \ ,
\quad \hat{\cal E}^{(U^{\dagger})}|\rho)=|U^{\dagger} \rho U). \]
If $UU^{\dagger}=U^{\dagger}U=I$, then
\[ \hat{\cal E}^{(U)}\hat{\cal E}^{(U^{\dagger})}=
\hat{\cal E}^{(U^{\dagger})}\hat{\cal E}^{(U)}=\hat I. \]
In the matrix representation we have
\[ \sum^{N-1}_{\alpha=0} {\cal E}^{(U)}_{\mu \alpha}
{\cal E}^{(U^{\dagger})}_{\alpha \nu}=
\sum^{N-1}_{\alpha=0} {\cal E}^{(U^{\dagger})}_{\mu \alpha}
{\cal E}^{(U)}_{\alpha \nu}=\delta_{\mu \nu}  \ , \]
i.e. ${\cal E}^{(U^{\dagger})}{\cal E}^{(U)}=
{\cal E}^{(U)}{\cal E}^{(U^{\dagger})}=I$.
Note that
\[ {\cal E}^{(U^{\dagger})}_{\mu \nu}=
\frac{1}{2^{n}}Tr\Bigl( \sigma_{\mu} U^{\dagger}\sigma_{\nu} U \Bigr)=
\frac{1}{2^{n}}Tr\Bigl( \sigma_{\nu} U \sigma_{\mu} U^{\dagger} \Bigr)=
{\cal E}^{(U)}_{\nu \mu}, \]
i.e. ${\cal E}^{(U^{\dagger})}=({\cal E}^{(U)})^{T}$.
Finally, we obtain (\ref{ORT}). \ \ \  $\Box$ \\

\bp
{\it If $\hat{\cal E}^{\dagger}$ is an adjoint superoperator for
linear trace-preserving superoperator $\hat{\cal E}$, then matrices
of the superoperator are connected by transposition
${\cal E}^{\dagger}={\cal E}^T$:}
\begin{equation}
({\cal E}^{\dagger})_{\mu \nu}={\cal E}_{\nu \mu}. \end{equation}
\ep

\noindent {\bf Proof.}
Using
\[ \hat{\cal E}=\sum^{m}_{j=1} \hat L_{A_{j}} \hat R_{A^{\dagger}_{j}} \ ,
\quad \hat{\cal E}^{\dagger}=\sum^{m}_{j=1} \hat L_{A^{\dagger}_{j}}
\hat R_{A_{j}}, \]
we get
\[ {\cal E}_{\mu \nu}=\frac{1}{2^{n}} \sum^{m}_{j=1}
Tr (\sigma_{\mu}A_{j}\sigma_{\nu}A^{\dagger}_{j}), \]
\[ ({\cal E}^{\dagger})_{\mu \nu}=\frac{1}{2^{n}} \sum^{m}_{j=1}
Tr (\sigma_{\mu}A^{\dagger}_{j}\sigma_{\nu}A_{j})=\]
\[=\frac{1}{2^{n}} \sum^{m}_{j=1} Tr
(\sigma_{\nu}A_{j}\sigma_{\mu}A^{\dagger}_{j})={\cal E}_{\nu \mu}. \]
Obviously, if we define the superoperator $\hat{\cal E}$ by
\[ \hat{\cal E}=\sum^{N-1}_{\mu=0}\sum^{N-1}_{\nu=0}
{\cal E}_{\mu \nu}|\mu)(\nu|, \]
then the adjoint superoperator has the form
\[ \hat{\cal E}^{\dagger}=\sum^{N-1}_{\mu=0}
\sum^{N-1}_{\nu=0}{\cal E}_{\nu \mu}|\mu)(\nu|. \]

\bp
{\it If $\hat{\cal E}^{\dagger} \hat{\cal E}=
\hat{\cal E} \hat{\cal E}^{\dagger}= \hat I$,
then $\hat{\cal E}$ is an orthogonal quantum four-valued logic gate, i.e.
${\cal E}^T{\cal E}={\cal E}{\cal E}^T=I$.}
\ep

\noindent {\bf Proof.}
If $\hat{\cal E}^{\dagger} \hat{\cal E}=\hat I$, then
\[ \sum^{N-1}_{\alpha=0} (\mu|\hat{\cal E}^{\dagger}|\alpha)(\alpha|
\hat{\cal E}|\nu)=(\mu|\hat I|\nu), \]
i.e.
\[ \sum^{N-1}_{\alpha=0} ({\cal E}^{\dagger})_{\mu \alpha}
{\cal E}_{\alpha \nu}=\delta_{\mu \nu}. \]
Using proposition 8 we have
\[ \sum^{N-1}_{\alpha=0} ({\cal E}^{T})_{\mu \alpha}
{\cal E}_{\alpha \nu}=\delta_{\mu \nu}, \]
i.e. ${\cal E}^{T}{\cal E}=I$. \ \ \  $\Box$ \\

Note that n-qubit unitary two-valued logic gate
$U$ is an element of Lie group $SU(2^{n})$.
The dimension of this group is equal to
$dim \ SU(2^{n})=(2^{n})^{2}-1=4^{n}-1$.
The matrix of n-ququat orthogonal linear gate
$\hat{\cal U}=\hat{\cal E}^{(U)}$ can be considered as 
an element of Lie group $SO(4^{n}-1)$.
The dimension of this group is equal to
$dim \ SO(4^{n}-1)=(4^{n}-1)(2 \cdot 4^{n-1}-1)$.

For example, if $n=1$, then
\[ dim \ SU(2^{1})=3 \ , \quad dim \  SO(4^{1}-1)=3. \]
If $n=2$, then
\[ dim \ SU(2^{2})=15 \ , \quad dim \  SO(4^{2}-1)=105. \]
Therefore, not all orthogonal quantum four-valued logic gates
for mixed and pure states are connected
with unitary two-valued logic gates for pure states.

\subsection{Single ququat orthogonal gates}

Let us consider single ququat quantum four-valued logic gate $\hat{\cal U}$
associated with unitary single qubit two-valued logic gate $U$.

\bp {\it Any single qubit unitary quantum two-valued logic gate
can be realized as the product of single ququat simple rotation
quantum four-valued logic gates
$\hat{\cal U}^{(1)}(\alpha)$, $\hat{\cal U}^{(2)}(\theta)$
and $\hat{\cal U}^{(1)}(\beta)$ defined by
\[ \hat{\cal U}^{(1)}(\alpha)=|0)(0|+|3)(3|+
\cos \alpha \Bigl(|1)(1|+|2)(2| \Bigr)+\]
\[+\sin \alpha \Bigl(|2)(1|-|1)(2| \Bigr), \]
\[ \hat{\cal U}^{(2)}(\theta)=|0)(0|+|2)(2|+
\cos \theta \Bigl(|1)(1|+|3)(3| \Bigr)+\]
\[+\sin \theta \Bigl(|1)(3|-|3)(1| \Bigr), \]
where $\alpha$, $\theta$ and $\beta$ are Euler angles.
}
\ep

\noindent {\bf Proof.}
Let us consider a general single qubit unitary gate \cite{Bar}.
Every unitary one-qubit gate $U$ can be represented by a 
$2 \times 2$-matrix
\[ U(\alpha,\theta,\beta)=e^{-i \alpha \sigma_{3}/2}
e^{-i \theta \sigma_{2}/2}e^{-i \beta\sigma_{3}/2}=
U_1(\alpha)U_2(\theta)U_1(\beta), \]
where
$$U_1(\alpha)=\left(
\begin{array}{ll}
e^{-i \alpha/2} & 0 \\
0        & e^{i \alpha/2}
\end{array}
\right),$$
$$U_2(\theta)=\left(
\begin{array}{rr}
\cos \theta/2 &- \sin \theta/2 \\
\sin \theta/2 & \cos \theta/2
\end{array}
\right),$$
where $\alpha$, $\theta$ and $\beta$ are Euler angles.
The corresponding $4\times 4$-matrix ${\cal U}(\alpha,\theta,\beta)$
of a four-valued logic gate has the form
\[ {\cal U}(\alpha,\theta,\beta)={\cal U}^{(1)}(\alpha)
{\cal U}^{(2)}(\theta){\cal U}^{(1)}(\beta), \]
where
\[ {\cal U}^{(1)}_{\mu \nu}(\alpha)=
\frac{1}{2}Tr\Bigl(\sigma_{\mu}
U_1(\alpha)\sigma_{\nu} U^{\dagger}_1(\alpha)\Bigr), \]
\[ {\cal U}^{(2)}_{\mu \nu}(\theta)=
\frac{1}{2}Tr\Bigl(\sigma_{\mu}
U_2(\theta)\sigma_{\nu} U^{\dagger}_2(\theta)\Bigr). \]
Finally, we obtain
$${\cal U}^{(1)}(\alpha)=\left(
\begin{array}{cccc}
1&0&0&0\\
0&\cos \alpha & -\sin \alpha & 0 \\
0&\sin \alpha & \cos \alpha & 0 \\
0&0 & 0 &  1
\end{array}
\right),$$
$${\cal U}^{(2)}(\theta)=\left(
\begin{array}{cccc}
1&0&0&0\\
0&\cos \theta & 0 & \sin \theta \\
0&0 & 1 & 0 \\
0&-\sin \theta & 0 & \cos \theta
\end{array}
\right),$$
where
\[ 0  \le \alpha <2 \pi , \quad  0\le \theta \le \pi ,
\quad 0 \le \beta\le 2\pi. \]
Using $U(\alpha,\theta +2\pi, \beta)=- U(\alpha,\theta, \beta)$,
we get that two-valued logic gates $U(\alpha,\theta, \beta)$ and
$U(\alpha,\theta +2\pi, \beta)$ map to single quantum
four-valued logic gate ${\cal U}(\alpha,\theta, \beta)$.
The back rotation four-valued logic gate is defined by the matrix
\[ {\cal U}^{-1}(\alpha,\theta, \beta)=
{\cal U}(2\pi -\alpha, \pi- \theta,2\pi - \beta) . \]
The simple rotation gates
$\hat{\cal U}^{(1)}(\alpha)$, $\hat{\cal U}^{(2)}(\theta)$,
$\hat{\cal U}^{(1)}(\beta)$
are defined by matrices $\hat{\cal U}^{(1)}(\alpha)$,
$\hat{\cal U}^{(2)}(\theta)$ and $\hat{\cal U}^{(1)}(\beta)$. \ \ \ $\Box$\\

Let us introduce simple reflection gates by
\[ \hat{\cal R}^{(1)}=|0)(0|-|1)(1|+|2)(2|+|3)(3|, \]
\[ \hat{\cal R}^{(2)}=|0)(0|+|1)(1|-|2)(2|+|3)(3|, \]
\[ \hat{\cal R}^{(3)}=|0)(0|+|1)(1|+|2)(2|-|3)(3|. \]

\bp
{\it Any single ququat linear quantum four-valued logic
gate $\hat{\cal E}$ defined by orthogonal matrix
${\cal E}:$ \ ${\cal E}{\cal E}^T=I$ can be realized by
\begin{itemize}
\item simple rotation gates $\hat{\cal U}^{(1)}$ and
$\hat{\cal U}^{(2)}$.
\item inversion gate $\hat {\cal I}$ defined by
\[ \hat{\cal I}=|0)(0|-|1)(1|-|2)(2|-|3)(3|. \]
\end{itemize}
}
\ep

\noindent {\bf Proof.} Using proposition 10 and
\[ \hat{\cal R}^{(3)}=\hat{\cal U}^{(1)} \hat{\cal I}, \quad
\hat{\cal R}^{(2)}=\hat{\cal U}^{(2)} \hat{\cal I} \ , \quad
\hat{\cal R}^{(1)}=\hat{\cal U}^{(1)}
\hat{\cal U}^{(1)} \hat{\cal I}, \]
we get this proposition. \ \ \ $\Box$\\

\noindent{\bf Example 1.}
In the generalized computational basis the Pauli matrices
as two-valued logic gates are the quantum four-valued logic gates
with diagonal $4\times 4$ matrices. The gate $I=\sigma_0$ is
\[ \hat{\cal U}^{(\sigma 0)}=\sum^3_{\mu=0}|\mu)(\mu|=\hat I, \]
i.e. \
${\cal U}^{(\sigma 0)}_{\mu \nu}=
(1/2)Tr(\sigma_{\mu}\sigma_{\nu})=\delta_{\mu \nu}$.

For the unitary quantum two-valued logic gates which are equal
to the Pauli matrix $\sigma_k$, where $k \in \{1,2,3\}$,
we have quantum four-valued logic gates
\[ \hat{\cal U}^{(\sigma k)}=\sum^3_{\mu , \nu=0}
{\cal U}^{(\sigma k)}_{\mu \nu} \ |\mu)(\nu|, \]
with the matrix
\begin{equation} \label{9}
{\cal U}^{(\sigma k)}_{\mu \nu}=2 \delta_{\mu 0} \delta_{\nu 0}+
2 \delta_{\mu k} \delta_{\nu k}- \delta_{\mu \nu} .
\end{equation}

\noindent{\bf Example 2.}
In the generalized computational basis the unitary NOT gate
("negation") of two-valued logic
\[ X=|0><1|+|1><0|=\sigma_{1}=
\left(
\begin{array}{cc}
0&1\\
1&0\\
\end{array}
\right), \]
is represented by the quantum four-valued logic gate
\[ \hat{\cal U}^{(X)}=|0)(0|+|1)(1|-|2)(2|-|3)(3|. \]

\noindent{\bf Example 3.}
The Hadamar two-valued logic gate
\[ H=\frac{1}{\sqrt{2}}(\sigma_1+\sigma_3)  \]
can be represented as a quantum four-valued logic gate by
\[ \hat{\cal E}^{(H)}=|0)(0|-|2)(2|+|3)(1|+|1)(3|, \]
with the matrix
\[ {\cal E}^{(H)}_{\mu \nu}=\delta_{\mu 0} \delta_{\nu 0}-
\delta_{\mu 2} \delta_{\nu 2}+\delta_{\mu 3} \delta_{\nu 1}+
\delta_{\mu 1} \delta_{\nu 3}. \]

\subsection{Measurements as quantum four-valued logic gates}

It is known that the von Neumann measurement
superoperator $\hat{\cal E}$ is defined by
\begin{equation} \label{vNm}
\hat {\cal E}|\rho)=\sum^{r}_{k=1}| P_{k}\rho P_{k}) \ , \end{equation}
where $\{P_{k}|k=1,..,r\}$ is a (not necessarily complete)
sequence of orthogonal projection operators on ${\cal H}^{(n)}$.

Let $P_{k}$ be projectors onto the pure state $|k>$
which define usual computational basis $\{|k>\}$, i.e.
\[ P_{k}=|k><k|. \]

\bp
{\it A nonlinear quantum four-valued logic gate $\hat{\cal N}$
for von Neumann measurement (\ref{vNm})
of the state $\rho=\sum^{N-1}_{\alpha=0}|\alpha) \rho_{\alpha}$
is defined by
\begin{equation} \hat{\cal N}=\sum^{r}_{k=1} \frac{1}{p(r)}
\sum^{N-1}_{\mu=0} \sum^{N-1}_{\nu=0}
{\cal E}^{(k)}_{\mu \nu} |\mu)(\nu|, \end{equation}
where
\begin{equation} \label{EE1}
{\cal E}^{(k)}_{\mu \nu}=
\frac{1}{2^{n}} Tr(\sigma_{\mu} P_{k} \sigma_{\nu} P_{k}),
\quad p(r)=\sqrt{2^n}\sum^r_{k=1}
\sum^{N-1}_{\alpha=0} {\cal E}^{(k)}_{0 \alpha}
\rho_{\alpha} . \end{equation}
}
\ep

\noindent {\bf Proof.}
The trace-decreasing superoperator $\hat {\cal E}_{k}$ is defined by
\[ |\rho) \ \rightarrow \ |\rho^{\prime})=
\hat {\cal E}_{k}|\rho)=|P_{k}\rho P_{k}). \]
This superoperator has
the form $\hat{\cal E}_{k}=\hat L_{P_{k}} \hat R_{P_{k}}$.
Then
\[ {\rho'}_{\mu}=(\mu|\rho')=(\mu|\hat{\cal E}_{k}|\rho)=
\sum^{N-1}_{\nu=0} (\mu|\hat{\cal E}_{k}|\nu)(\nu|\rho)=
\sum^{N-1}_{\nu=0} {\cal E}^{(k)}_{\mu \nu} \rho_{\nu} , \]
where
\[ {\cal E}^{(k)}_{\mu \nu}=(\mu|\hat{\cal E}_{k}|\nu)=
\frac{1}{2^{n}} Tr(\sigma_{\mu} P_{k} \sigma_{\nu} P_{k}). \]
The probability that the process represented by $\hat{\cal E}_{k}$
occurs is
\[ p(k)=Tr(\hat{\cal E}_k(\rho))=(I|\hat{\cal E}_{k}|\rho)=
\sqrt{2^n}\rho^{\prime}_0=\sqrt{2^n}
\sum^{N-1}_{\alpha=0} {\cal E}^{(k)}_{0 \alpha} \rho_{\alpha} . \]
If
\[ p(r)=\sqrt{2^n} \sum^r_{k=1} \sum^{N-1}_{\alpha=0}
{\cal E}^{(k)}_{0 \alpha} \rho_{\alpha} \not=0 , \]
then the matrix for nonlinear trace-preserving gate $\hat{\cal N}$ is
\[ {\cal N}_{\mu \nu}=\sum^{r}_{k=1} p^{-1}(r)
{\cal E}^{(k)}_{\mu \nu}. \]

\noindent{\bf Example.}
Let us consider single ququat projection operator
\[ P_{0}=|0><0|=\frac{1}{2}(\sigma_{0}+\sigma_{3}). \]
Using formula (\ref{EE1}) we derive
\[ {\cal E}^{(0)}_{\mu \nu}=
\frac{1}{8}Tr\Bigl(\sigma_{\mu}(\sigma_{0}+\sigma_{3})
\sigma_{\nu}(\sigma_{0}+\sigma_{3}) \Bigr)=\]
\[=\frac{1}{2}\Bigl(
\delta_{\mu 0}\delta_{\nu 0}+\delta_{\mu 3} \delta_{\nu 3}+
\delta_{\mu 3}\delta_{\nu 0}+\delta_{\mu 0} \delta_{\nu 3} \Bigr). \]

The linear trace-decreasing superoperator
for von Neumann measurement
projector $|0><0|$ onto pure state $|0>$ is
\[ \hat{\cal E}^{(0)}=\frac{1}{2}\Bigl( |0)(0|+|3)(3|+
|0)(3|+|3)(0| \Bigr). \]

\noindent{\bf Example.}
For the projection operator
\[ P_{1}=|1><1|=\frac{1}{2}(\sigma_{0}-\sigma_{3}) \]
Using formula (\ref{EE1}) we derive
\[ {\cal E}^{(1)}_{\mu \nu}=
\frac{1}{2}\Bigl(
\delta_{\mu 0}\delta_{\nu 0}+\delta_{\mu 3} \delta_{\nu 3}-
\delta_{\mu 3}\delta_{\nu 0}-\delta_{\mu 0} \delta_{\nu 3}
\Bigr). \]
The linear superoperator $\hat{\cal E}^{(1)}$ for 
the von Neumann measurement projector onto the pure state $|1>$ is
\[ \hat{\cal E}^{(1)}=
\frac{1}{2}\Bigl( |0)(0|+|3)(3|-|0)(3|-|3)(0| \Bigr). \]

The superoperators $\hat{\cal E}^{(0)}$ and $\hat{\cal E}^{(1)}$
are not trace-preserving.
The probabilities that processes represented by superoperators
$\hat{\cal E}^{(k)}$ occurs are
\[ p(0)=\frac{1}{\sqrt{2}}(\rho_0+\rho_{3}) \ , \quad
p(1)=\frac{1}{\sqrt{2}}(\rho_0-\rho_{3}). \]


\section{Classical four-valued logic gates}


Let us consider some elements of classical four-valued logic.
For the concept of many-valued logic, see \cite{Re,RT,Ya1,Ya2,Ta}.

\subsection{Elementary classical gates}

A  function $g(x_{1},...,x_{n})$ describes a
classical four-valued logic gate if the following conditions hold:
\begin{itemize}
\item all $x_{i} \in \{0,1,2,3\}$, where $i=1,...,n$.
\item $g(x_{1},...,x_{n}) \in \{0,1,2,3\}$.
\end{itemize}

It is known that
the number of all classical logic gates with n-arguments
$x_{1},...,x_{n}$ is equal to $4^{4^{n}}$.
The number of  classical logic gates $g(x)$ with single argument
is equal to $4^{4^{1}}=256$. Let us write some of these gates in the table.

\vskip 5mm
\begin{tabular}{|c|c|c|c|c|c|c|c|c|}
\hline
\multicolumn{9}{|c|}{Single argument classical gates}\\
\hline
x& $\sim x$ &$\Box x$&$\diamondsuit x$&$\overline{x}$&$I_{0}$&$I_{1}$&$I_{2}$&$I_{3}$\\ \hline
0& 3        &   0    &      0         &      1       &  3    &   0   &   0   & 0   \\
1& 2        &   0    &      3         &      2       &  0    &   3   &   0   & 0    \\
2& 1        &   0    &      3         &      3       &  0    &   0   &   3   & 0   \\
3& 0        &   3    &      3         &      0       &  0    &   0   &   0   & 3    \\ \hline
\end{tabular}
\vskip 3mm

The number of classical logic gates $g(x_{1},x_{2})$ with
two-arguments is equal to
\[ 4^{4^{2}}=4^{16}=42949677296. \]
Let us write some of these classical gates in the table.

\vskip 5mm
\begin{tabular}{|c|c|c|c|c|}
\hline
\multicolumn{5}{|c|}{Two-arguments classical gates}\\
\hline
$(x_{1},x_{2})$&$\land$&$\lor$&$V_{4}$&$\sim V_{4}$ \\ \hline
(0;0)          &   0   &   0  &  1    &    2        \\
(0;1)          &   0   &   1  &  2    &    1        \\
(0;2)          &   0   &   2  &  3    &    0        \\
(0;3)          &   0   &   3  &  0    &    3        \\
(1;0)          &   0   &   1  &  2    &    1        \\
(1;1)          &   1   &   1  &  2    &    1        \\
(1;2)          &   1   &   2  &  3    &    0        \\
(1;3)          &   1   &   3  &  0    &    3        \\
(2;0)          &   0   &   2  &  3    &    0        \\
(2;1)          &   1   &   2  &  3    &    0        \\
(2;2)          &   2   &   2  &  3    &    0        \\
(2;3)          &   2   &   3  &  0    &    3        \\
(3;0)          &   0   &   3  &  0    &    3        \\
(3;1)          &   1   &   3  &  0    &    3        \\
(3;2)          &   2   &   3  &  0    &    3        \\
(3;3)          &   3   &   3  &  0    &    3        \\ \hline
\end{tabular}
\vskip 3mm

Let us define some elementary classical four-valued logic gates by formulae.
\begin{itemize}
\item Luckasiewicz negation: \  $\sim x=3-x$.
\item Cyclic shift: \ $\overline{x}=x+1(mod4)$.
\item Functions $I_{i}(x)$, where $i=0,...,3$, such that
$I_{i}(x)=3$ if $x=i$ and $I_{i}(x)=0$ if $x\not=i$.
\item Generalized conjunction: \ $x_{1} \land x_{2}=min(x_{1},x_{2})$.
\item Generalized disjunction: \ $x_{1} \lor x_{2}=max(x_{1},x_{2})$.
\item Generalized Sheffer-Webb function: \
$$V_{4}(x_{1},x_{2})=max(x_{1},x_{2})+1(mod 4).$$
\end{itemize}

The generalized conjunction and disjunction satisfy the
commutative law, associative law and distributive law:
\begin{itemize}
\item Commutative law
\[ x_{1} \land x_{2}=x_{2} \land x_{1} \ , \quad
x_{1} \lor x_{2}=x_{2} \lor x_{1} \]
\item Associative law
\[ (x_{1} \lor x_{2}) \lor x_{3}=
x_{1} \lor (x_{2} \lor x_{3}). \]
\[ (x_{1} \land x_{2}) \land x_{3}=
x_{1} \land (x_{2} \land x_{3}). \]
\item Distributive law
\[ x_{1}\lor (x_{2} \land x_{3})=
(x_{1} \lor x_{2}) \land ( x_{1} \lor x_{3}). \]
\[ x_{1}\land (x_{2} \lor x_{3})=
(x_{1} \land x_{2}) \lor ( x_{1} \land x_{3}). \]
\end{itemize}
Note that the Luckasiewicz negation satisfies the following properties: 
\[ \sim(\sim x)=x \ , \quad
\sim(x_{1} \land x_{2})=(\sim x_{1}) \lor (\sim x_{2}). \]
The following usual negation rules are not satisfied by
the circular shift: 
\[ \overline{\overline{x}} \not=x \ , \quad \overline{x_{1}\land x_{2}}
\not=\overline{x_{1}} \lor \overline{x_{2}}. \]
The analog of the 
disjunction normal form of the n-argument classical
four-valued logic gate is
\[ g(x_{1},...,x_{n})=\bigvee_{(k_{1},...,k_{n})} \ I_{k_{1}}(x_{1})
\land ... \land I_{k_{n}}(x_{n}) \land g(k_{1},..,k_{n}). \]

Let us consider (functional) complete sets \cite{Ya1,Ya2} of
classical four-valued logic gates.

\bp
{\it The set $\{0, 1, 2, 3, I_{0}, I_{1}, I_{2}, I_{3},
x_{1} \land x_{2}, x_{1} \lor x_{2}\}$ is a complete set.\\
The set $\{\overline{x}, x_{1} \lor x_{2}\}$ is a complete set.\\
The gate $V_{4}(x_{1}, x_{2})$ is a complete set.
}
\ep

\noindent {\bf Proof.} This proposition is proved in \cite{Ya2}. 
\ \ \ $\Box$

\subsection{Quantum gates for single argument classical gates}

Let us consider linear trace-preserving quantum gates for
classical gates $\sim$, $\overline{x}$,
$I_{0},I_{1},I_{2},I_{3}$, $0,1,2,3$,
$\diamondsuit$, $\Box$.

\bp
{\it Any single argument classical four-valued logic gate $g(\nu)$
can be realized as a linear trace-preserving quantum
four-valued logic gate by
\[ \hat{\cal E}^{(g)}=|0)(0|+\sum^{3}_{k=1 }|g(k))(k|+\]
\begin{equation} +(1-\delta_{0g(0) })\Bigl( |g(0))(0|-
\sum^{3}_{\mu=0} \sum^{3}_{\nu=0}(1-\delta_{\mu g(\nu)})
|\mu)(\nu| \Bigr). \end{equation}
}
\ep

\noindent {\bf Proof.}
The proof is by direct calculation in
\[ \hat{\cal E}^{(g)}|\alpha]=|g(\alpha)], \]
where
\[ \hat{\cal E}^{(g)}|\alpha]= \frac{1}{\sqrt{2}}\Bigl(
\hat{\cal E}^{(g)}|0)+\hat{\cal E}^{(g)}|\alpha) \Bigr). \]

\noindent{\bf Examples.}\\
1. The Luckasiewicz negation gate is
\[ \hat{\cal E}^{(\sim)}=|0)(0|+|1)(2|+|2)(1|+|3)(0|-|3)(3|. \]
2.  The four-valued logic gate $I_{0}$ can be realized by
\[ \hat{\cal E}^{(I_0)}=|0)(0|+|3)(0|-\sum^{3}_{k=1}|3)(k|. \]
3. The gates $I_{k}(x)$, where $k=1,2,3$ are
\[ \hat{\cal E}^{(I_k)}=|0)(0|+|3)(k|. \]
4.  The gate $\overline{x}$ can be realized by
\[ \hat{\cal E}^{(\overline{x})}=|0)(0|+|1)(0|+|2)(1|+|3)(2|-
\sum^{3}_{k=1}|1)(k|. \]
5. The constant gates $0$ and $k=1,2,3$ can be realized by
\[ \hat{\cal E}^{(0)}=|0)(0|  \ , \quad
\hat{\cal E}^{(k)}=|0)(0|+|k)(0|. \]
6.  The gate $\diamondsuit x$ is realized by
\[ \hat{\cal E}^{(\diamondsuit)}=|0)(0|+\sum^{3}_{k=1}|3)(k|. \]
7. The gate $\Box x= \sim \diamondsuit x$ is
\[ \hat{\cal E}^{(\Box)}=|0)(0|+|3)(3|. \]
Note that the quantum four-valued logic gates $\hat{\cal E}^{(\sim)}$,
$\hat{\cal E}^{(I_{0})}$, $\hat{\cal E}^{(k)}$,
$\hat{\cal E}^{(g_{1})}$ are not unital gates.

\subsection{Quantum gates for two-arguments classical gates}

Let us consider quantum four-valued logic gates for
two-arguments classical four-valued logic gates.

1. The generalized conjunction $x_{1} \land x_{2}=min(x_{1},x_{2})$ and
generalized disjunction $x_{1} \land x_{2}=max(x_{1},x_{2})$ can be
realized by a two-ququat quantum four-valued logic gate with $T=0$:
\[ \hat{\cal E}|x_1,x_2]=|x_{1} \lor x_{2},x_{1} \land x_{2}] . \]
Let us write the quantum four-valued logic gate
which realizes the gate in the generalized computational basis by
\[ \hat{\cal E}=\sum^{N-1}_{\mu=0}  \sum^{N-1}_{\nu=0} |\mu\nu)(\mu \nu|+
\sum^{3}_{k=1}\Bigl( |0k)-|k0) \Bigr)(k 0|+ \]
\[ +\sum^{3}_{k=2}\Bigl( |1k)-|k1) \Bigr)(k 1|+
\Bigl( |23)-|32) \Bigl)(32|. \]

2.  The Sheffer-Webb function gate $|x_{1},x_{2}] \  \rightarrow \
|V_{4}(x_{1},x_{2}),\sim V_{4}(x_{1},x_{2})]$ can be realized
by a two-ququat quantum gate with $T\not=0$:
\[ \hat{\cal E}^{(SW)}=|00)(00|+|12)(00|-
\sum^{3}_{\mu=0}\sum^{3}_{\nu=1}|12)(\mu \nu|+|21)(10|+ \]
\[ +|21)(11|+|30)(02|+|30)(20|+|30)(12|+|30)(21|+\]
\[ +|30)(22|+|03)(03|+|03)(13|+|03)(23|+\sum^{3}_{\mu=0} |03)(3\mu|. \]
Note that this Sheffer-Webb function gate is not a unital quantum gate and
\[ \hat{\cal E}^{(SW)} \not=
|V_{4}(x_{1},x_{2}),\sim V_{4}(x_{1},x_{2}))(x_{1},x_{2}|. \]


\section{Universal set of quantum four-valued logic gates }


The condition for performing arbitrary unitary operations to
realize a quantum computation by dynamics of a closed quantum
system is well understood  \cite{Bar,DBE,DV,LL}.
Using quantum unitary gates, a quantum computer with pure states
may realize the time sequence of operations corresponding to any
unitary dynamics. Deutsch, Barenco and Ekert \cite{DBE},
DiVincenzo \cite{DV} and Lloyd \cite{LL}
showed that almost any two-qubit quantum unitary gate
is universal for a quantum computer with pure states. It is
known \cite{Bar,DBE,DV,LL} that a set of quantum gates that
consists of all one-qubit unitary gates and the two-qubit exclusive-OR
(XOR) gate is universal for quantum computer with pure states
in the sense that all unitary operations
on arbitrary many qubits can be expressed as compositions of these
unitary gates. Recently in \cite{BB} universality for a quantum computer 
with n-qudits quantum unitary gates on pure states was considered.

The same is not true for the general quantum operations
(superoperators) corresponding to the dynamics of open quantum systems.
In the paper \cite{Bac} a single qubit
open quantum system with Markovian dynamics was considered and
the resources needed for universality of general quantum
operations were studied.
An analysis of completely positive trace-preserving superoperators on
single qubit density matrices was realized in papers \cite{FA,KR,RSW}.

Let us study universality for quantum four-valued logic
gates \cite{Tarpr,Tarpr2}. \\

\noindent {\bf Definition}
{\it A set of quantum four-valued logic gates is universal iff all
quantum gates on arbitrary many ququats can be expressed
as compositions of these gates.}\\

Single ququat gates cannot map two initially
un-entangled ququats into an entangled state.
Therefore, the single ququat gates or set of single ququat gates
are not universal gates for a quantum computer with mixed states.
Quantum gates which are realizations of classical gates
cannot be universal by definition, since these gates evolve
generalized computational states to generalized computational states
and never to the superposition of them.

The matrix ${\cal E}$ of the linear real superoperator $\hat{\cal E}$ 
on $\overline{\cal H}^{(n)}$ is 
an element of Lie group $TGL(4^{n}-1,\mathbb{R})$.
The linear superoperator $\hat{\cal E}$ on
$\overline{\cal H}^{(n)}$ is a quantum four-valued logic gate 
(completely positive trace-preserving superoperator) iff 
the matrix ${\cal E}$ is a completely positive element
of Lie group $TGL(4^{n}-1,\mathbb{R})$.
The matrix ${\cal N}$ of a nonlinear real trace-preserving
superoperator $\hat {\cal N}$ on $\overline{\cal H}^{(n)}$
is a quantum four-valued logic gate defined by 
\begin{equation} \label{NNN}
\hat{\cal N}(\rho)=
\frac{\hat{\cal E}(\rho)}{Tr(\hat{\cal E}(\rho))} \end{equation}
iff the matrix ${\cal E}$ of the linear trace-decreasing 
superoperator $\hat{\cal E}$ is a completely positive element
of Lie group $GL(4^{n},{\bf R})$.
The condition of complete positivity leads to difficult
inequalities for matrix elements \cite{Choi,FA,KR,RSW}. 
In order to satisfy the condition of complete positivity we use 
the following representation:   
\begin{equation} \label{ELR} \hat{\cal E}=\sum^{m}_{j=1} \hat L_{A_{j}}
\hat R_{A^{\dagger}_{j}} \ , \end{equation}
where $\hat L_{A}$ and $\hat R_{A}$ are left and right multiplication
superoperators on $\overline{\cal H}^{(n)}$ defined by
$\hat L_{A}|B)=|AB)$, $\hat R_{A}|B)=|BA)$.
It is known that any linear completely positive 
superoperator $\hat{\cal E}$ can be represented by (\ref{ELR}).
Any trace-decreasing superoperator (\ref{ELR})
generates a quantum four-valued logic gate by (\ref{NNN}).
{\it To find the universal set of
completely positive (linear or nonlinear) superoperators, i.e.
quantum four-valued logic gates, we suggest
considering the universal set of the superoperators 
$\hat L_{A_j}$ and $\hat R_{A^{\dagger}_j}$.}
Let the superoperators $\hat L_{A_j}$ and $\hat R_{A^{\dagger}_j}$
be called "pseudo-gates".
A set of pseudo-gates is universal iff all
pseudo-gates on arbitrary many ququats can be expressed
as compositions of these pseudo-gates.
The matrices of the superoperators $\hat L_{A}$ and $\hat R_{A^{\dagger}}$
are connected by complex conjugation.
The set of these matrices is a group $GL(4^n,\mathbb{C})$.
Obviously, the universal set of
pseudo-gates $\hat L_{A}$ is connected with a universal set of completely
positive superoperators $\hat{\cal E}$ of the quantum
four-valued logic gates.

The trace-preserving condition for linear superoperator (\ref{ELR})
is equivalent to the requirement ${\cal E}_{0 \mu}=\delta_{0 \mu}$
for gate matrix ${\cal E}$.
The trace-decreasing condition can be satisfied by inequality
of the following proposition.

\bp {\it If the matrix elements ${\cal E}_{\mu \nu}$ of a
superoperator $\hat{\cal E}$ are satisfied by the inequality
\begin{equation} \label{ieq1}
\sum^{N-1}_{\mu=0} ({\cal E}_{0 \mu})^{2} \le 1 , \end{equation}
then  $\hat{\cal E}$ is a trace-decreasing superoperator.}
\ep

\noindent {\bf Proof.}
Using the Schwarz inequality
\[ \Bigl( \sum^{N-1}_{\mu=0} {\cal E}_{0 \mu} \rho_{\mu} \Bigr)^{2}\le
\sum^{N-1}_{\mu=0} ({\cal E}_{0 \mu})^{2}
\sum^{N-1}_{\nu=0} (\rho_{\nu})^{2}  \]
and the property of the density matrix
\[ Tr \rho^{2}=(\rho|\rho)=\sum^{N-1}_{\nu=0} (\rho_{\nu})^{2} \le 1, \]
we have
\[ |Tr\hat{\cal E}(\rho)|^{2}= |(0|\hat{\cal E}|\rho)|^{2}=
\Bigl(\sum^{N-1}_{\mu=0} {\cal E}_{0 \mu} \rho_{\mu} \Bigr)^{2}
\le \sum^{N-1}_{\mu=0} ({\cal E}_{0 \mu})^{2}. \]
Using (\ref{ieq1}), we get $|Tr \hat{\cal E}(\rho)| \le 1$.
Since $\hat{\cal E}$ is a completely positive (or positive)
superoperator ($\hat{\cal E}(\rho) \ge 0$), it follows that
\[ 0 \le Tr \hat{\cal E}(\rho) \le 1, \]
i.e. $\hat{\cal E}$ is a trace-decreasing superoperator. \ \ \ $\Box$ \\

Let us consider the superoperators $\hat L_{A}$ and $\hat R_{A^{\dagger}}$.
These superoperators can be represented by
\begin{equation} \label{LARA}
\hat L_A=  \sum^{N-1}_{\mu=0}\sum^{N-1}_{\nu=0}
L^{(A)}_{\mu \nu} |\mu)(\nu|, \quad
\hat R_{A^{\dagger}}= \sum^{N-1}_{\mu=0}  \sum^{N-1}_{\nu=0}
R^{(A^{\dagger})}_{\mu \nu} |\mu)(\nu|, \end{equation}
where matrices
$L^{(A)}_{\mu \nu}$ and $R^{(A^{\dagger})}_{\mu \nu}$ are defined by
\[ L^{(A)}_{\mu \nu}=
\frac{1}{2^n} Tr\Bigl(\sigma_{\mu} A \sigma_{\nu} \Bigr)=
\frac{1}{2^n} Tr\Bigl(\sigma_{\alpha} \sigma_{\mu} A  \Bigr), \]
\[ R^{(A^{\dagger})}_{\mu \nu}=\frac{1}{2^n}
Tr\Bigl(\sigma_{\mu} \sigma_{\nu} A^{\dagger} \Bigr)
=\frac{1}{2^n}
Tr\Bigl(A^{\dagger}\sigma_{\mu} \sigma_{\nu} \Bigr). \]

\bp
{\it The matrix ${\cal E}_{\mu \nu}$ of the completely positive
superoperator (\ref{ELR}) can be represented by}
\begin{equation} \label{ELR-M} {\cal E}_{\mu \nu}
=\sum^{m}_{j=1} \sum^{N-1}_{\alpha=0} L^{(jA)}_{\mu \alpha}
R^{(jA^{\dagger})}_{\alpha \nu} . \end{equation}
\ep

\noindent {\bf Proof.}
Let us write the matrix ${\cal E}_{\mu \nu}$ by matrices of
superoperators $\hat L_{A_j}$ and $\hat R_{A_j}$.
\[ {\cal E}_{\mu \nu}=(\mu|\hat{\cal E} |\nu)=
\sum^{m}_{j=1} (\mu| \hat L_{A_j}
\hat R_{A^{\dagger}_j} |\nu)=\]
\[=\sum^{m}_{j=1} \sum^{N-1}_{\alpha=0} (\mu| \hat L_{A_j}|\alpha)
(\alpha|\hat R_{A^{\dagger}_j} |\nu)=
\sum^{m}_{j=1} \sum^{N-1}_{\alpha=0}  L^{(jA)}_{\mu \alpha}
R^{(jA^{\dagger})}_{\alpha \nu} . \]
Finally, we obtain (\ref{ELR-M}),
where
\[ L^{(jA)}_{\mu \alpha}=(\mu|\hat L_{A}|\alpha)=
\frac{1}{2^n} (\sigma_{\mu}| \hat L_{A_j}| \sigma_{\alpha})=\]
\[=\frac{1}{2^n} Tr\Bigl(\sigma_{\mu} A_j \sigma_{\alpha} \Bigr)=
\frac{1}{2^n} Tr\Bigl(\sigma_{\alpha} \sigma_{\mu} A_j  \Bigr), \]
\[ R^{(jA^{\dagger})}_{\alpha \nu}=(\alpha|\hat R_{A^{\dagger}_j}|\nu)=
\frac{1}{2^n} (\sigma_{\alpha}| \hat R_{A^{\dagger}_j}|\sigma_{\nu})=\]
\[=\frac{1}{2^n}
Tr\Bigl(\sigma_{\alpha} \sigma_{\nu} A^{\dagger}_j \Bigr)
=\frac{1}{2^n}
Tr\Bigl(A^{\dagger}_j\sigma_{\alpha} \sigma_{\nu} \Bigr). \]
The matrix elements can be rewritten in the form
\begin{equation} \label{LRmat} L^{(jA)}_{\mu \alpha}=\frac{1}{2^n}
(\sigma_{\mu} \sigma_{\alpha}| A_j) \ , \quad
R^{(jA^{\dagger})}_{\alpha \nu}=
\frac{1}{2^n}(A_j|\sigma_{\alpha} \sigma_\nu). \end{equation}

\noindent{\bf Example.}
Let us consider the single ququat pseudo-gate $\hat L_{A}$.
The elements of pseudo-gate matrix $L^{(A)}$ are defined by
\[ L^{(A)}_{\mu \nu}=\frac{1}{2}Tr(\sigma_{\mu} A \sigma_{\nu}). \]
Let us denote
\[ a_{\mu}=\frac{1}{2}Tr(\sigma_{\mu} A). \]
Using
\[ L^{(A)}_{kl}=\frac{1}{2}Tr(\sigma_{l} \sigma_{k} A)=
\frac{1}{2}\delta_{kl}TrA +
\frac{i}{2}\varepsilon_{lkm} Tr(\sigma_{m} A), \]
where $k,l,m=1,2,3$, we get
\[ \hat L_{A}=\sum^{3}_{\mu=0} a_{0} |\mu)(\mu|+
\sum^{3}_{k=0} a_{k} \Bigl( |0)(k|+ |k)(0|\Bigr)+ \]
\[ +ia_{1}\Bigl( |3)(2|-|2)(3| \Bigr)+
ia_{2}\Bigl( |1)(3|-|3)(1| \Bigr)+\]
\[+ia_{3}\Bigl( |2)(1|-|1)(2| \Bigr). \]
The pseudo-gate matrix is
\[ L^{(A)}_{\mu \nu}= \delta_{\mu \nu}Tr A+
\sum^{3}_{m=1}\Bigl( \delta_{\mu 0}\delta_{\nu m} +
\delta_{\mu m}\delta_{\nu 0}\Bigr) Tr(\sigma_{m}A)+ \]
\begin{equation} +i\sum^{3}_{m=1} \delta_{\mu k}\delta_{\nu l}
\varepsilon_{lkm}  Tr(\sigma_{m}A). \end{equation}

Let us consider the properties of the matrix elements
$L^{(jA)}_{\mu \alpha}$ and $R^{(jA^{\dagger})}_{\mu \alpha}$.

\bp
{\it The matrices
$L^{(jA)}_{\mu \alpha}$ and $R^{(jA^{\dagger})}_{\mu \alpha}$
are complex $4^{n}\times 4^{n}$ matrices and their elements
are connected by complex conjugation: }
\begin{equation} (L^{(jA)}_{\mu \alpha})^{*}=
R^{(jA^{\dagger})}_{\mu \alpha}. \end{equation}
\ep

\noindent {\bf Proof.}
Using complex conjugation of the matrix elements (\ref{LRmat}), we get
\[ (L^{(jA)}_{\mu \alpha})^{*}= \frac{1}{2^n}
(\sigma_{\mu} \sigma_{\alpha}| A_j)^{*}=
\frac{1}{2^n} (A_j|\sigma_{\mu} \sigma_{\alpha})=
R^{(jA^{\dagger})}_{\mu \alpha}. \]

We can write the gate matrix (\ref{ELR-M}) in the form
\begin{equation} {\cal E}_{\mu \nu}
=\sum^{m}_{j=1} \sum^{N-1}_{\alpha=0} L^{(jA)}_{\mu \alpha}
(L^{(jA)}_{\alpha \nu})^{*}. \end{equation}


\bp
{\it The matrices $L^{(jA)}_{\mu \alpha}$ and
$R^{(jA^{\dagger})}_{\mu \alpha}$ of the n-ququat
quantum four-valued logic gate (\ref{ELR}) are the elements
of Lie group $GL(4^{n},\mathbb{C})$.
The set of these matrices is a group.}
\ep

\noindent {\bf Proof.} The proof is trivial. \ \ \ $\Box$\\

A superoperator $\hat{\cal E}$ on $\overline{\cal H}^{(2)}$ is called
primitive \cite{BB} if $\hat{\cal E}$ maps the tensor product of 
single ququats to the tensor product of single ququats, i.e.
if $|\rho_{1})$ and $|\rho_{2})$ are ququats, then
we can find ququats $|\rho^{\prime}_{1})$ and $|\rho^{\prime}_{2})$
such that
\[ \hat{\cal E}|\rho_{1} \otimes\rho_{2})=
|\rho^{\prime}_{1} \otimes \rho^{\prime}_{2}). \]
The superoperator $ \hat{\cal E}$ is called imprimitive if
$\hat{\cal E}$ is not primitive.

It can be shown that almost every pseudo-gate that operates
on two or more ququats is a universal pseudo-gate.

\bp
{\it The set of all single ququat pseudo-gates and
any imprimitive two-ququat pseudo-gate
are universal set of pseudo-gates.}
\ep


\noindent {\bf Proof.} This proposition can be proved by analogy
with \cite{DV,DBE,BB}. Let us consider some points of the proof.
Expressed in  group theory language, all n-ququat pseudo-gates
are elements of the Lie group $GL(4^n,\mathbb{C})$. Two-ququat
pseudo-gates $\hat L$ are elements of Lie group $GL(16,\mathbb{C})$.
The question of universality is the same as that of what
set of superoperators $\hat L$ is sufficient to generate $GL(16,\mathbb{C})$.
The group $GL(16,\mathbb{C})$ has $(16)^2=256$
independent one-parameter subgroups $GL_{\mu \nu}(16,\mathbb{C})$
of one-parameter pseudo-gates $\hat L^{(\mu \nu)}(t)$ such that
$\hat L^{(\mu \nu)}(t)=t|\mu)(\nu|$.
Infinitesimal generators of Lie group $GL(4^n,\mathbb{C})$ are defined by
\begin{equation} \hat H_{\mu \nu}=\Bigl(\frac{d}{dt}
\hat L^{(\mu \nu)}(t) \Bigr)_{t=0}, \end{equation}
where $\mu,\nu=0,1,...,4^{n}-1$.
The generators $\hat H_{\mu \nu}$ of the one-parameter subgroup
$GL_{\mu \nu}(4^n,\mathbb{R})$ are superoperators of the form
$\hat H_{\mu \nu}=|\mu)(\nu|$ on $\overline{\cal H}^{(n)}$
which can be represented by
$4^n \times 4^n$ matrices $H_{\mu \nu}$ with elements
\[ (H_{\mu \nu})_{\alpha \beta}=\delta_{\alpha \mu} \delta_{\beta \nu}. \]
The set of superoperators $\hat H_{\mu \nu}$ is a basis
(Weyl basis \cite{BR})
of Lie algebra $gl(16,\mathbb{R})$ such that
\[ [\hat H_{\mu \nu},\hat H_{\alpha \beta}]=
\delta_{\nu \alpha} \hat H_{\mu \beta}- \delta_{\mu \beta} \hat
H_{\nu \alpha}, \]
where $\mu, \nu, \alpha, \beta =0,1,...,15.$
Any element $\hat H$ of the algebra
$gl(16,\mathbb{C})$ can be represented by
\[ \hat H=\sum^{15}_{\mu=0}\sum^{15}_{\nu=0}
h_{\mu \nu} \hat H_{\mu \nu}, \]
where $h_{\mu \nu}$ are complex coefficients.

As a basis of Lie algebra $gl(16,\mathbb{C})$
we can use $256$ linearly independent
self-adjoint superoperators
\[ H_{\alpha \alpha}=|\alpha)(\alpha|, \quad
H^r_{\alpha \beta}=|\alpha)(\beta|+|\beta)(\alpha|,\]
\[ H^i_{\alpha \beta}= -i\Bigl(
|\alpha)(\beta|-|\beta)(\alpha|\Bigr), \]
where $0\le \alpha \le \beta \le 15$.
The matrices of these generators are
Hermitian $16 \times 16$ matrices.
The matrix elements of 256 Hermitian $16 \times 16$ matrices
$H_{\alpha \alpha}$,  $H^r_{\alpha \beta}$ and $H^{i}_{\alpha \beta}$
are defined by
\[ (H_{\alpha \alpha})_{\mu \nu}=
\delta_{\mu \alpha} \delta_{\nu \alpha} \ ,
\quad (H^r_{\alpha \beta})_{\mu \nu}=
\delta_{\mu \alpha} \delta_{\nu \beta}
+\delta_{\mu \beta} \delta_{\nu \alpha}, \]
\[ (H^i_{\alpha \beta})_{\mu \nu}=
-i(\delta_{\mu \alpha} \delta_{\nu \beta}
-\delta_{\mu \beta} \delta_{\nu \alpha}). \]
For any Hermitian generator $\hat H$ there exists
a one-parameter pseudo-gate $\hat L(t)$
which can be represented in the form
$\hat L(t)=exp \ it \hat H$ such that
$\hat L^{\dagger}(t)\hat L(t)=\hat I$.

Let us write the main operations which allow us to derive
new pseudo-gates $\hat L$ from a set of pseudo-gates.\\
\begin{enumerate}
\item We introduce general SWAP (twist)
pseudo-gate $\hat T^{(SW)}$.
A new pseudo-gate $\hat L^{(SW)}$ defined by
$\hat L^{(SW)}=\hat T^{(SW)} \hat L \hat T^{(SW)}$
is obtained directly from $\hat L$ by exchanging two ququats.
\item Any superoperator $\hat L$ on $\overline{\cal H}^{(2)}$
generated by the commutator
$i[\hat H_{\mu\nu}, \hat H_{\alpha \beta}]$ can be obtained
from $\hat L_{\mu\nu}(t)=exp \ it\hat H_{\mu\nu}$
and $\hat L_{\alpha \beta}(t)=exp \ it\hat H_{\alpha \beta}$ because
\[ exp \ t \ [\hat H_{\mu\nu},\hat H_{\alpha \beta}]=\]
\[=\lim_{n \rightarrow \infty} \Bigl( \hat L_{\alpha \beta}(-t_n)
\hat L_{\mu\nu}(t_n) \hat L_{\alpha \beta}(t_n) \hat
L_{\mu\nu}(-t_n)\Bigr)^n, \]
where $t_n=1/\sqrt{n}$.
Thus we can use the commutator
$i[\hat H_{\mu \nu}, \hat H_{\alpha \beta}]$
to generate pseudo-gates.
\item Every transformation $\hat L(a,b)=exp i\hat H(a,b)$
of $GL(16,\mathbb{C})$ generated by
superoperator $\hat H(a,b)=a\hat H_{\mu\nu}+b\hat H_{\alpha \beta}$, where
$a$ and $b$ are complex, can obtained from
$\hat L_{\mu\nu}(t)=exp \ it\hat H_{\mu\nu}$
and $\hat L_{\alpha \beta}(t)=exp \ it\hat H_{\alpha \beta}$   by
 \[ exp \ i \hat H(a,b)=
\lim_{n \rightarrow \infty} \Bigl( \hat L_{\mu \nu}(\frac{a}{n})
\hat L_{\alpha \beta}(\frac{b}{n})\Bigr)^n. \]
\end{enumerate}

For other details of the proof, see \cite{DV,DBE,BB}
and \cite{Bare,Bar,LL}.

\section{Conclusion }

In this paper we demonstrate a model of quantum computations
with mixed states. The computations are realized by quantum
operations, not necessarily unitary. Mixed states subject
to the general quantum operations could increase efficiency.
This increase is connected with the increasing number of
computational basis elements for Hilbert space.
A pure state of n two level quantum systems is an element
of $2^n$-dimensional functional Hilbert space.
A mixed state of the system is an element of $(2^n)^2=4^n$-dimensional
operator Hilbert space. The conventional quantum two-valued logic
is replaced by quantum four-valued logic.
Therefore the increased efficiency can be formalized in terms of a
four-valued logic replacing the conventional two-valued logic.
Unitary gates and quantum operations for a quantum computer with pure states
and two-valued logic can be considered as four-valued logic gates
of a mixed-state quantum computer. Quantum algorithms \cite{Shor1,Shor2,Grover}
on a quantum computer with mixed states are expected to run
on a smaller network than with pure state implementation.

In the quantum computer model with pure states, control of quantum
unitary gates is realized by classical parameters of the Hamilton operator.
Open and closed quantum systems can be described by the generalized
von Neumann equation \cite{Tarpr,kn1,kn2}:
\begin{equation} \label{vNe} \frac{\partial}{\partial t} \rho(t)=
\hat \Lambda \rho(t), \end{equation}
where $\hat \Lambda$ is the Liouville superoperator.
For closed quantum systems this superoperator is defined by Hamiltonian $H$:
\[ \hat \Lambda=-\frac{i}{\hbar}(\hat L_H-\hat R_H), \]
where $\hat L_H$ and $\hat R_H$ are superoperators defined by
$\hat L_{H}\rho=H \rho$ and $\hat R_{H}\rho=\rho H$.
Quantum unitary gates on pure states are controlled by classical
parameters entering the Hamiltonian $H$.
For open quantum systems with completely positive evolution
the Liouville superoperator $\hat \Lambda$ is given by
\[ \hat \Lambda=-\frac{i}{\hbar}(\hat L_{H}-\hat R_{H})+
\frac{1}{2\hbar}\sum^{m}_{j=1}\Bigl(
2\hat L_{V_{j}} \hat R_{V^{\dagger}_{j}}-
\hat L_{V^{ }_{j}} \hat L_{V^{\dagger}_{j}}-
\hat R_{V^{\dagger}_{j}} \hat R_{V^{ }_{j}} \Bigr), \]
where $H$ is a bounded self-adjoint Hamilton operator,
$\{V_j\}$ is a sequence of bounded operators
\cite{Lin,GKS,GFVKS,PZ,AL,Tarpr,kn1,kn2}.
Quantum four-valued logic gates on mixed states are controlled
by classical parameters of the Hamiltonian $H$ and the bounded
operators $V_j$ \ \cite{LV,kn2}.

In the paper we consider universality for
general quantum four-valued logic gates acting on mixed states.
The matrices of the quantum gates can be considered
as elements of some matrix group but these matrices are completely
positive (or positive) elements of this matrix group.
The condition of complete positivity leads to difficult
inequalities for matrix elements \cite{Choi,FA,KR,RSW}. 
The completely positive condition for quantum four-valued logic gates 
can be satisfied by Kraus representation (\ref{ELR}).
To find the universal set of quantum four-valued logic gates 
we suggest considering the universal set of the superoperators
(\ref{LARA}) called pseudo-gates.
Pseudo-gates are not necessarily completely positive and the
set of pseudo-gates matrices is a group.
In the paper we show that almost any two-ququat pseudo-gate is universal.

In the usual quantum computer model a measurement of the final pure state
is described by projection operators $P_k=|k><k|$.
In the suggested model a measurement of the final mixed state 
can be described by projection superoperators \cite{FS} described by
$\hat {\cal P}_{\mu}=|\mu)(\mu|$,
where $|\mu)$ are defined by (\ref{mu}) and (\ref{mu2}).

A scenario for laboratory realization of quantum computations by
quantum operations with mixed states can be a generalization
of the scheme \cite{BP}. The quantum gates on mixed states can
be realized by controlled polarization of the laser field.
The control of the field polarization leads to control
of the polarization mixed state of the electron.
The scheme can use polarization sensitive optical fluorescence
and single photon detection for read-out.


\end{multicols}{2}


\end{document}